# ARTICLE

# Multifunctional photoresponsive organic molecule for electric field sensing and modulation

Yingmu Zhang[a], Jinghan He[b], Patrick J.G. Saris[a], Hyun Uk Chae[c], Subrata Das[c], Rehan Kapadia[c], Andrea M. Armani [a b c] *

Nonlinear optical organic molecules have advanced a wide range of fields spanning from integrated photonics to biological imaging. With advances in molecular design, an emerging application is multifunctional nonlinear organic materials. Unlike conventional molecules which simply emit light through single or multi photon processes, multifunctional materials can perform multiple tasks, such as modulating the optical signal or reporting an electric field intensity. In this work, we report a multifunctional organic molecular device with electric field 'sense-and-modulate' capability. This is achieved by combining two distinct functional modules. The electric field-reporting module relies on a photo-induced electron transfer (PeT) dye, tetraphenylethylene (TPE) as a two-photon (2p) imaging agent; while the electric field-modulating module relies on the organic photoconductor, naphthalimide (NAI). To reduce cross-talk between the two modules, they are separated by a long alkyl chain. The photophysical properties and photoconductivity of the probe molecule are studied in a range of solvents and in solid state, and the results agree with the density functional theory predictions. Specifically, 2p excitation is demonstrated, and the photoconductivity is rapid and reversible. The entire system is optically controlled, including signal read-out, and the two modules can be operated simultaneously or individually.

## Introduction

Whether organic or inorganic, fluorescent molecules and particles have been a critical tool of discovery and an innovation platform for nearly 200 years. For example, fluorophores have revealed cell structure and function, been used as indicators for detecting toxins, and have served as the active layer in many optoelectronic devices.[1–6] This prior work inspired the concept of an organic molecular device, or a device comprised of a single or cluster of organic molecules. Initial research in this area leveraged existing probe molecules to demonstrate biodetection (eg molecular beacons) and biomolecular motors based on optical or mechanical responses.[7–11] Given their lengthscale, the potential applications of molecular devices extend far beyond biological systems into optoelectronics and quantum devices. For example, recent work has explored stimulated emission materials, organic solid-state lasers, and frequency-upconversion.[12–18] However, to create a molecular device capable of performing multiple functions requires combining several functional groups into a single molecular construct.

One of the key challenges in designing multifunctional molecules is reducing the cross-talk between the different excitation mechanisms. This type of interference can reduce the overall system efficiency and performance.[19,20] With recent advances in machine learning and computational design algorithms, material chemists are able to optimize molecular designs accelerating experimental efforts.[21–23]

In the present work, we develop and characterize an all-optical multifunctional molecular device designed to simultaneously sense and modulate electric fields. The molecule, named as NAI-TPE-PyS, is comprised of two non-interacting modules that are tethered together by a non-interacting spacer (Fig. 1). The sensor module is derived from tetraphenylethylene (TPE), which is a two-photon (2p)

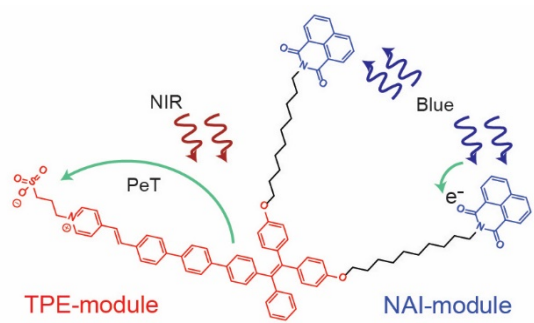

Fig. 1. Chemical structure of NAI-TPE-PyS with TPE-module in red and NAI-module in blue.

[a.] Mork Family Department of Chemical Engineering and Materials Science, University of Southern California, Los Angeles, California 90089, USA.
[b.] Department of Chemistry, University of Southern California, Los Angeles, California 90089, USA.
[c.] Ming Hsieh Department of Electrical and Computer Engineering-Electrophysics, University of Southern California, Los Angeles, California 90089, USA

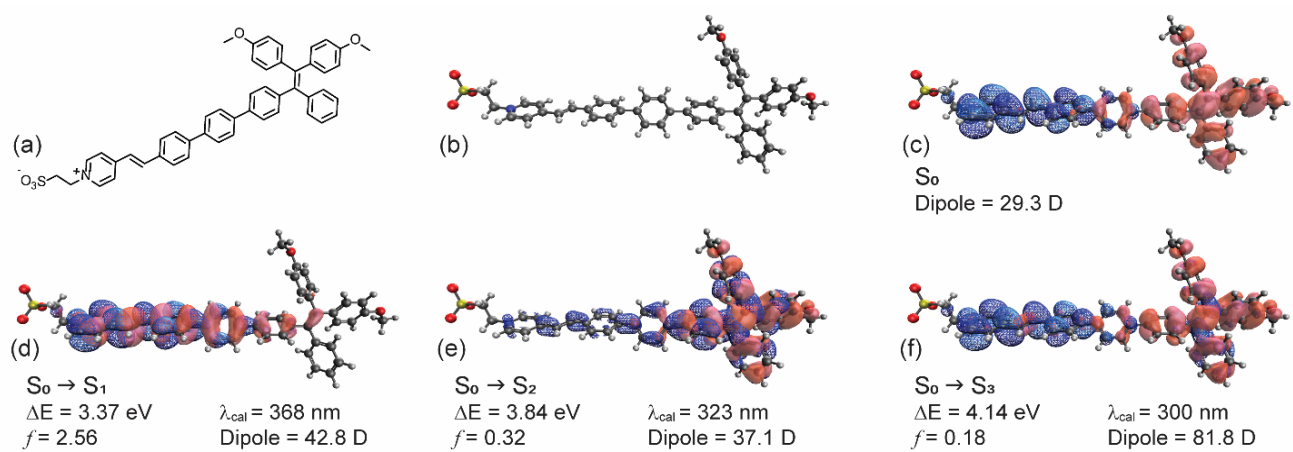

Fig. 2. TPE-PyS model compound: a) chemical structure; b) optimized ground state geometry; c) HOMO (red solid) and LUMO (blue mesh) iso-surfaces (0.05 Å$^{-3}$); natural transition orbital iso-surfaces (0.05 Å$^{-3}$) for hole (red solid) and electron (blue mesh) wavefunctions: d) S$_1$, e) S$_2$, and f) S$_3$.

fluorophore that is excited in the near-IR (NIR). The modulator module is an organic photoconductor, naphthalimide (NAI), whose resistivity can be tuned using an ultraviolet optical source. The non-interacting alkyl chain does not affect the photophysical properties of the modules but rather reduces Dexter energy transfer between modules by physical spacing.

The TPE and NAI modules were judiciously chosen for low overlap of their absorption and emission spectra to reduce Foerster resonant energy transfer. However, the precise chemical structure of the molecule is optimized using density functional theory as part of this work. The tether enforces colocalization of the modules to reduce the possibility for bulk phase separation of the modules, which could occur for a simple blend of the two components under the processing conditions for various applications. The entire system is optically controlled, including signal read-out, and the two modules can be operated simultaneously or individually (Fig.1).

The TPE sensor module relies on the ultrafast photo-induced electron transfer (PET) process to allow large changes in the electric field to be detected. Unlike voltage sensitive dyes[24–28], PeT molecules show superb signal-to-noise ratio, sensitivity, and response speed.[29–33] This performance is inherent to the ultrafast PeT process, which is an intramolecular electron transfer through its donor-spacer-acceptor backbone upon exposure to excitation sources, competing with the radiative fluorescent emission.[34,35] In addition, TPE exhibits a unique behavior known as aggregation-induced emission (AIE). In an AIE material, the emission signal in the condensed state is increased due to restricted intramolecular rotation (RIR).[36,37] Among PET and AIE materials, TPE is unique because it exhibits both behaviors.

As mentioned, NAI is a photoconductor.[38–41] This type of molecule allows the resistivity to be modulated with light. Although organic photoconductors have been widely used in optoelectronics,[42–47] their integration into a multifunctional molecule is underexplored.[48] Moreover, an organic molecular device with both electric field sensor and modulator capabilities has yet to be demonstrated.

## Results and discussions
### Molecular Design and Computational Modelling

The molecular device is comprised of two covalently coupled modules: a tetraphenylene (TPE)-based PeT dye connected with naphthalimide (NAI)-based photoconductor.

As mentioned, the NAI "modulate" module was chosen due to its extensive use as an organic photoconductor and was incorporated without structural modification. The "sense" module of NAI-TPE-PyS was designed based on TPE but was modified with PyS to create TPE-PyS, representing the donor-spacer-acceptor system, as required by a PeT dye.[49]

To assist in this design, the reporter part was modelled in *silico* based on a truncated structure, namely, Me-TPE-PyS (Fig. 2a & 2b). The alkyl linkers and peryleneimide moieties of NAI-TPE-PyS are separated from electronic communication with TPE-module and therefore were replaced with methyl groups to reduce the computational cost of the modelling. The ground state equilibrium geometry and frontier orbital densities of Me-TPE-PyS were calculated by density functional theory (DFT) in the gas phase at the B3LYP/6-31g* level of theory. With TPE as an electron donating group and PyS as an electron accepting one, the HOMO density resides on the TPE moiety with some extension onto the phenylene linker (Fig. 2c, red solid). The LUMO density resides on the pyridinium and benzylidene moieties with some extension onto the phenylene linker (Fig. 2c, blue mesh).

The excited states were calculated by TD-DFT with a conductor-like polarizable continuum medium (dielectric = 78.39, water) at the CAM-B3LYP/6-311++G** level of theory. Fig. 2d-f shows the natural transition orbitals for the lowest three possible singlet states along with corresponding transition properties. Accordingly, the first singlet excited state (S$_1$) has a highly allowed vertical transition with an energy of 3.37 eV (368 nm) (Fig. 2d). The dipole moment of S$_1$ (42.8 D) is 18.9 D more than that of the ground state (29.3 D), suggesting that it may be moderately solvatochromic. This transition is characterized by a π-π* transition on the pyridinium and phenylene linker, corresponding to electron transfer from lower lying occupied orbitals to the LUMO. The second singlet excited state ($S_2$) has a vertical transition energy of 3.84 eV (323 nm) (Fig. 2e). The dipole moment of $S_2$ (37.1 D) is 7.8 D more than that of the ground state (29.3 D), suggesting that it may be weakly solvatochromic. This transition is characterized by a localized π-π* transition on the TPE moiety, corresponding to electron transfer from the HOMO to higher lying unoccupied orbitals. The third singlet excited state ($S_3$) has a vertical transition energy of 4.14 eV (300 nm) (Fig. 2f). The dipole moment of $S_3$ (81.8 D) is 52.5 D more than that of the ground state (29.3 D), suggesting that it may be strongly solvatochromic. This transition corresponds to the HOMO to LUMO transition, representing charge transfer from TPE to PyS. Notably, based on this analysis, solvatochromicity, which is an indicator of voltage sensitivity, is anticipated.

### Synthesis

The synthetic route of NAI-TPE-PyS is summarized in scheme 1, and the synthesis details are included in the supplemental information (Scheme S1, Fig. S1–S14, ESI†). Using a Suzuki coupling reaction, a TPE unit, 4,4′-(2-(4-Bromophenyl)-2-phenylethene-1,1-diyl)-diphenol, was conjugated with a phenylenevinylene spacer at the bromine terminal to form HO-TPE-CHO. The hydroxide terminals were then connected to two alkylated NAI moieties via Williamson etherification. Finally, a Knoevenagel condensation of the aldehyde terminal with an electron-withdrawing pyridinium inner salt, 3-(4-methylpyridin-1-ium-1-yl)propane-1-sulfonate, yielded NAI-TPE-PyS as an orange powder with a 32% overall yield.

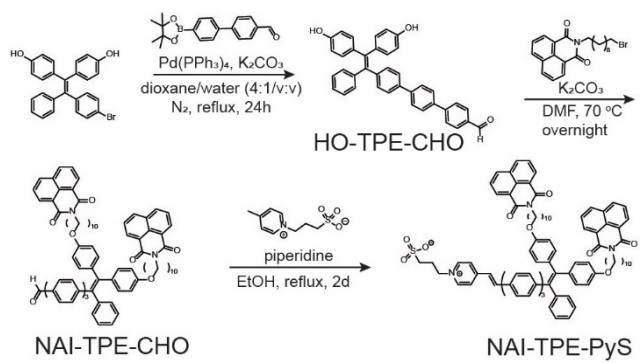

**Scheme 1.** The synthesis scheme of NAI-TPE-PyS.

### Spectroscopic Characterization

Due to the combination of the hydrophobic NAI terminals and the hydrophilic pyridinium inner salt terminal, NAI-TPE-PyS can be readily dissolved in polar aprotic solvents like DMSO and DMF as well as in low-polarity lipophilic solvents like DCM and chloroform. However, it shows low solubility in protic solvents like water or PBS buffer. Accordingly, the steady-state photophysical properties of NAI-TPE-PyS were analysed both in solutions with different polarity and in solid state thin films using ultraviolet-visible (UV-vis) and photoluminescence (PL) spectroscopy.

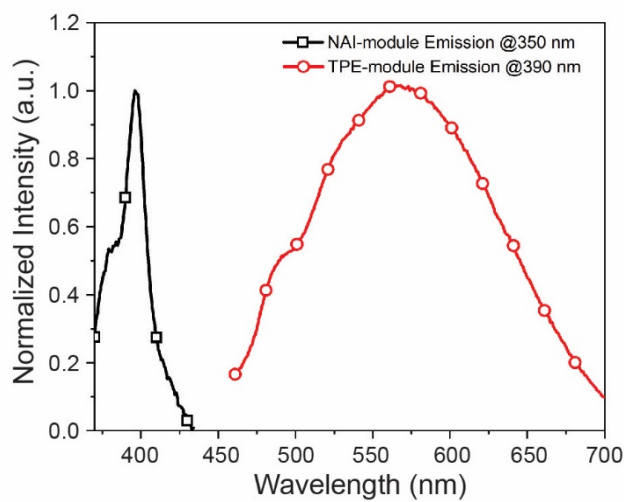

**Fig 3.** Emission spectra of NAI-TPE-PyS at λex = 350 nm (black squares) and λex= 390 nm (red circles) in PBS buffer (Concentration: 10 μM, containing 1% DMSO).

TPE is a moderate electron donor, and the pyridinium inner salt is a strong electron acceptor. Thus, the interaction between these compounds will lower the HOMO-LOMO energy gap, facilitating the intramolecular charge transfer (CT) process. As a result, the normalized absorption spectra of the molecule exhibit a broad profile tailing to 500 nm (Fig. S15, ESI†). Detailed photophysical data of NAI-TPE-PyS in selective solvents are shown in Fig. S16 and in Table S1 (ESI†). At shorter wavelengths, around 300-380 nm, the absorption band can be assigned to the n-π* and π-π* transitions of NAI- and the elongated TPE-conjugation moieties which overlaps with the absorption of NAI-TPE-CHO precursor. The absorption band at wavelengths longer than 380 nm, which is not present in the NAI-TPE-CHO precursor, can be assigned to the CT transition from TPE to pyridinium inner salt.

The two modules of NAI-TPE-PyS can be activated under independent optical control (Fig. 3). When excited at 350 nm, NAI-TPE-PyS displays characteristic narrowed emission of naphthalene derivatives centered at 400 nm, which indicates that the emission mainly stems from NAI-based module. When excited at 390 nm, NAI-TPE-PyS emits orange luminescence with the maximum peak at 585 nm dominated by the emission of TPE-based PeT dye, while no emission of NAI-module was observed. Consequently, the dual-component compound can be controlled by two wavelengths, where 390 nm is used to excite the TPE-module for voltage imaging and 350 nm is used to drive the photo-conductive NAI-module for voltage modulation. The difference in energy between these wavelengths (circa 400 meV) is enough to prevent crosstalk between the two modules.

To investigate the predicted solvatochromism of the TPE-PyS module, the emission was characterized in several solvents. When excited at 390 nm in PBS, the NAI-TPE-PyS emitted at 582nm(Fig. 4). The photoluminescent quantum yield in an aqueous solution (10 μM, containing 1% DMSO) was measured to be 0.02 (2%). When the NAI-TPE-PyS was a solid, the quantum yield increased to 8%, providing preliminary evidence of AIE behavior.

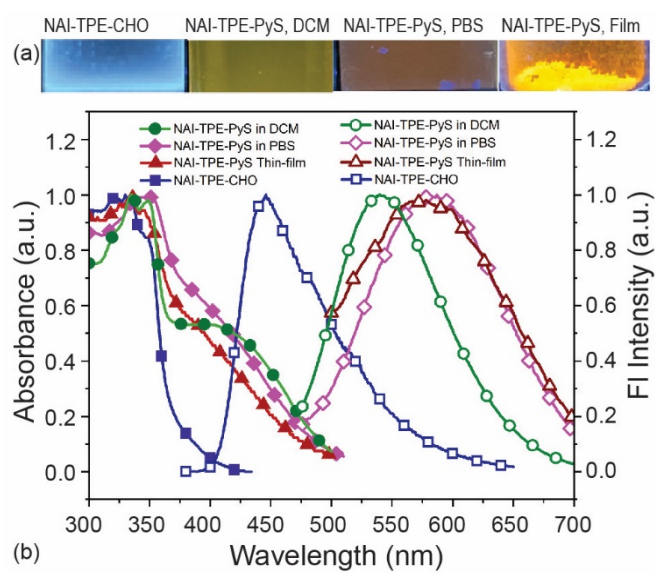

Fig. 4. Optical absorption and emission properties of NAI-TPE-PyS. (a) Images of fluorescence from NAI-TPE-CHO and NAI-TPE-PyS solutions and solid powder ($\lambda_{ex}$= 365 nm). (b) Normalized absorption spectra (solid symbols) and emission spectra (hollow symbols) of NAI-TPE-PyS in DCM or PBS or in solid state, and NAI-TPE-CHO at $\lambda_{ex}$= 390 nm.

This emission wavelength blue-shifts when the solvent is changed to DCM (Fig. 4). To further confirm this behavior, the emission was characterized over a range of solvents with varying polarities from chloroform to water, and a linear Lippert-Mataga correlation was established by plotting Stokes shifts ( $v_{abs}$–$v_{em}$ ) against solvent orientation polarizability ($\Delta f$ ) (Fig. S19, ESI†).[50] From the slope of the Lippert-Mataga plot, the increase of dipole moment ($\mu_e$-$\mu_g$) from ground state to excited state was calculated to be 29.8 D. Therefore, using the estimation of $\mu_e$ from the DFT model (59.1 D), the ground state dipole ($\mu_g$) is calculated to be 29.3 D. This large value indicates that the CT-component of the excited state will be stabilized in polar media.[51,52]

To further investigate the CT process, the related molecule NAI-TPE-CHO was synthesized and characterized. The maximum emission peak of this molecule was at 420 nm, representing a large bathochromic shift of 162 nm from the NAI-TPE-PyS compound. This difference can be ascribed to the donor-spacer-acceptor structure. Such interplay also explains the relatively weak emission of NAI-TPE-PyS compared to other published TPE-based AIEgens,[53] as non-radiative deactivation of the CT state competes with the radiative process.

Based on the combination of the computational and experimental results, we conclude that the reporter module will be a voltage sensitive dye. Computationally, the reporter has the proper ordering of excited state energies and dipole moments to facilitate voltage dependence. Spectroscopically, the synthesized dye displays the characteristic solvatochromism associated with voltage sensitive fluorophores due to sensing of the dielectric environment. We expect that it will demonstrate voltage sensitive luminescence when the molecule's motion is restricted, for example, if the molecule is inserted and aligned in a cellular membrane.

### Aggregation Induced Emission Properties

To rigorously investigate the AIE behavior of the TPE core, two approaches were used to initiate aggregation in a controlled manner. First, poor or low efficacy solvents were used. In our studies, DMSO was an ideal solvent for NAI-TPE-PyS, and THF was an antisolvent for NAI-TPE-PyS. Thus, by using a mixture of DMSO:THF, aggregates were induced. Second, the concentration of the NAI-TPE-PyS molecule in water was systematically increased, resulting in aggregation or micelle formation.

The DMSO solution of NAI-TPE-PyS (20 µM) exhibited weak fluorescence centered at 545 nm (Fig. S20, ESI†). As the concentration of THF was increased, the emission intensity increased, due to the formation of aggregates. Finally, at a THF fraction of 99%, the emission intensity of the mixture reached a maximum which was 3-fold higher than that of the initial DMSO solution (Fig. S20, ESI†). Based on the results in Fig. S20, the introduction of the alkyl NAI terminals did not eliminate the AIE response. However, it should be noted that, unlike other previously reported TPE-based AIEgens whose emission experiences significant enhancement upon aggregation in solution, the emission intensity of NAI-TPE-PyS at high THF fraction merely increases three times. This could be attributed to the long alkyl chains of the NAI-TPE-PyS interfering with the aggregation process, ultimately limiting the size of the aggregates.[51] Using dynamic light scattering (DLS), the diameter of the nanoaggregates was measured to be approximately 500 nm (Fig. S21, ESI†). Similar increases in emission intensity were observed as the concentration of NAI-TPE-PyS in water was increased from 0.01 to 80 µM (Fig. S20, ESI†).

### Nonlinear Optical Properties

In addition to single photon excitation processes, NAI-TPE-PyS can also undergo two-photon excitation processes. The emission decay lifetime in a DMSO solution under 750 nm excitation was measured using a confocal microscope-based Time-Correlated Single Photon Counting (TCSPC) setup. The TCSPC histogram and exponential decay fit are provided in Fig. S22 (ESI†). Fluorescence at wavelengths shorter than 550 nm were observed from NAI-TPE-PyS with a monoexponential excited state decay time of 160 ps. Using the photoluminescent quantum yield of 0.02 measured previously, the radiative and non-radiative rate constants are calculated to be 1.25 x $10^8$ $s^{-1}$ and 6.125 x $10^9$ $s^{-1}$ respectively, which is consistent with a weakly emitting organic fluorophore. These results confirm the predicted two-photon behavior of NAI-TPE-PyS using NIR excitation (750 nm). This ability is a critical stepping stone in the development of a two-photon voltage imaging and modulation probe.

### Photoconductivity Characterization

The 'modulate'-functionality relies on the NAI-module of the NAI-TPE-PyS. To characterize the photoconductivity, a two-terminal device was fabricated with Ti/Au electrodes, and a thin film of NAI-TPE-PyS was deposited (Fig. 5a, inset). Current vs. voltage measurements were performed both in dark and under light (350 nm) illumination. As shown in the I-V curves in Fig. 5a, the device was insulating in the dark. Upon illumination with 19mW of 350nm light, photo-induced current was observed clearly. This increase is attributed to the electron-hole pair generation and dissociation within the NAI-module. The stability and real-time change in photo-induced conductance of the device were further studied based on current vs. time measurements at a fixed voltage of 10 V. From the I-t curves in Fig. 5b, photo-induced current can be clearly distinguished from the device. The photocurrent increased or decreased sharply as the light was turned on and off, and the dark current was stable. The device also displayed an increase in responsivity as the optical power increased, demonstrating the power-dependent photoconduction behavior of NAI-module (Fig. 5c). It is noteworthy that the photocurrent remained stable over the course of these measurements which were performed in an ambient environment (Fig. S23, Table S2 ESI†).

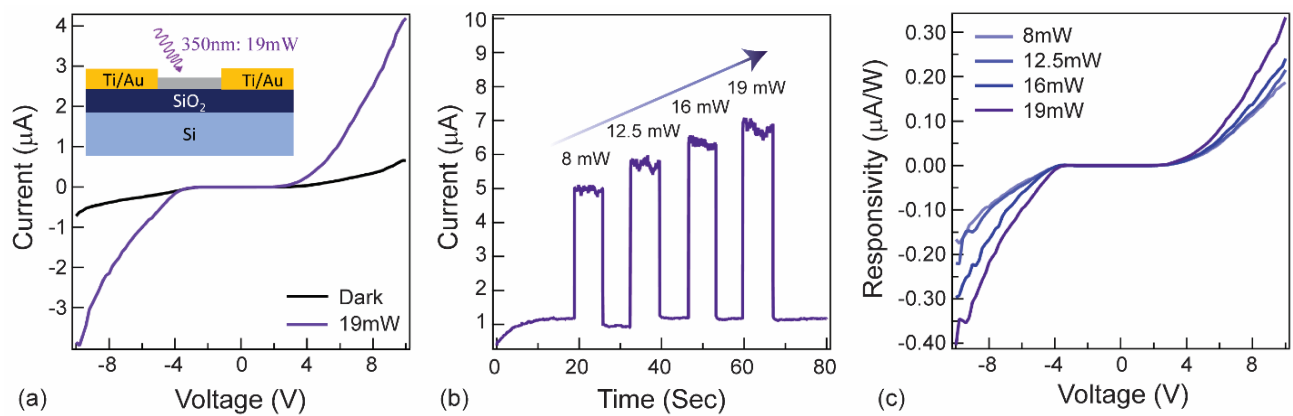

**Fig. 5.** Photoconductivity of NAI-TPE-PyS. (a) current vs. voltage measurements of the device. Inset: Schematic of NAI-TPE-PyS-coated two-terminal device. $SiO_2$: 2 μm; Ti: 5 nm; Au: 100 nm. Device channel size: Length*width = 200 μm*5000 μm. (b) current vs. time measurements of the device at different incident optical powers. (c) Responsivity vs. voltage of the device at different incident optical powers.

## Conclusions

In summary, the multifunctional molecular device, NAI-TPE-PyS, has been successfully designed, synthesized, and characterized. Electric field sensing and modulation abilities are derived from its two covalently coupled modules: a TPE-derived module connected by an alkyl chain to an NAI-derived module. To prove the potential of NAI-TPE-PyS as a multifunctional electric field molecular probe, the photophysical and optoelectronic properties have been investigated experimentally and theoretically.

The TPE-module is designed for voltage imaging where the TPE core is the donor and is conjugated with a pyridinium inner salt ($Py^+SO_3^-$) serving as the acceptor, forming the D-π-A backbone required by a PeT dye. The module exhibits a broad absorption profile and a solvent-dependent emission owing to the intramolecular charge transfer (CT) between the donor and acceptor. The CT-component of the excited state has been further confirmed through theoretical calculations. In addition, the module displays two-photon excited emission upon NIR irradiation. Photo-induced current is observed when the module is actuated with UV illumination, which results from hole-electron generation and transportation within the NAI moieties.

The study provides a new strategy for developing multifunctional molecules by synergistically combining the emerging fields of small molecule photoconductors and small molecule multi-photon fluorophores. In the future, the multifunctional molecule demonstrated here could prove to be a valuable tool in understanding the complex field of bioelectricity and in designing integrated optoelectronic quantum circuits.

## Experimental section

### Materials and Instruments

All commercially available chemical reagents and solvents were either purchased from VWR or Sigma Aldrich. Unless otherwise noted, they were used without further purification. $^1H$ and $^{13}C$-NMR spectra were recorded on a Varian Mercury 400 MHz spectrometer with 96-spinner sampler changer using either deuterated chloroform or DMSO as solvent, as indicated. UV-Vis absorption spectra were measured on Beckman Coulter Life Science UV/Vis spectrophotometer, DU 730 with a wavelength resolution of 2 nm. Steady-state fluorescence spectra were recorded on Horiba Scientific Fluoromax-4 spectrofluorometer with excitation slit width of 5 nm and emission slit width of 5 nm. Quantum yield was determined using a Quanta-φ integrating sphere attached to the Fluoromax-4. Time-resolved photoluminescence (TRPL) measurements were performed with a standard confocal microscope-based Time-Correlated Single Photon Counting (TCSPC) setup.

### Synthesis of HO-TPE-CHO

4'-(4,4,5,5-tetramethyl-1,3,2-dioxaborolan-2-yl)-[1,1'-biphenyl]-4-carbaldehyde (2.3 mmol, 0.7 g), 4,4'-(2-(4-Bromophenyl)-2-phenylethene-1,1-diyl)-diphenol (1.9 mmol, 0.84 g), $Pd(PPh_3)_4$ (0.19 mmol, 200 mg) and $K_2CO_3$ (7.6 mmol, 1.05 g) were added into a 100 mL round-bottom flask. The flask was fitted on the Schlenk line, vacuumed, and refilled with nitrogen iteratively three times. A mixing solvent (dioxane/water: 40 mL/10 mL) was bubbled with nitrogen for 30 min and then transferred to the flask through a canula. The mixture was then allowed to react for 24 hours at 100 °C. After cooling to the room temperature, the mixture was poured into water, and the pH was adjusted to 5. Then, the system was extracted with dichloromethane (DCM) and washed with water three times. The organic solvent was then removed and the crude solid recrystallized by hexane/ DCM to give the pure product. $^1$H NMR (400 MHz, Chloroform-d) δ 10.06 (s, 1H), 7.96 (d, J = 8.1 Hz, 2H), 7.79 (d, J = 8.4 Hz, 2H), 7.68 (s, 4H), 7.41 (d, J = 8.2 Hz, 3H), 7.12 (d, J = 2.2 Hz, 4H), 7.06 (s, 2H), 6.93 (dd, J = 16.6, 8.5 Hz, 4H), 6.59 (t, J = 9.0 Hz, 4H). $^{13}$C-NMR (101MHz, Chloroform-d): 191.91, 154.22, 154.14, 146.48, 144.10, 143.74, 140.85, 140.34, 138.79, 137.56, 136.40, 135.16, 132.78, 131.90, 131.42, 130.32, 127.75, 127.64, 127.46, 127.37, 126.22, 126.13, 114.68, 114.56.

**Synthesis of NAI-TPE-CHO**

Into a 250 mL two-necked round-bottom flask was added $K_2CO_3$ (248.8 mg, 1.8 mmol), 2-(10-Bromodecyl)-1H-benz[de]isoquinoline-1,3(2H)-dione (500 mg, 1.2 mmol) and HO-TPE-CHO (163 mg, 0.3 mmol). The flask was vacuumed and purged with dry $N_2$ three times. Then DMF (15 mL) was added, and the reaction was stirred overnight at 70 °C. After cooling to room temperature, the mixture was poured into water, extracted with DCM, washed with distilled water several times, and dried with anhydrous magnesium sulfate. The crude product was purified by silica column chromatography with hexane and ethyl acetate (gradient to 1:1/v:v) as eluent to give NAI-TPE-CHO as a yellow viscous oil. $^1$H NMR (400 MHz, Chloroform-d) δ 9.97 (s, 1H), 8.54 – 8.48 (m, 4H), 8.16 – 8.11 (m, 4H), 7.87 (d, J = 8.4 Hz, 2H), 7.74 – 7.63 (m, 7H), 7.60 (s, 4H), 7.32 (d, J = 8.4 Hz, 2H), 7.07 – 6.97 (m, 6H), 6.87 (dd, J = 17.3, 8.8 Hz, 4H), 6.60 – 6.51 (m, 4H), 4.12 – 4.07 (m, 4H), 3.79 (t, J = 6.5 Hz, 4H), 1.71 – 1.59 (m, 8H), 1.28 (d, J = 42.0 Hz, 24H). $^{13}$C-NMR (101MHz, Chloroform-d): 191.52, 164.19, 133.83, 132.61, 132.21, 131.56, 131.16, 130.30, 128.13, 127.79, 127.60, 127.42, 127.34, 126.09, 122.73, 113.90, 68.23, 40.48, 32.75, 32.64, 29.38, 29.32, 29.29, 29.28, 28.08, 28.04, 27.09, 26.02, 25.95, 25.78, 25.64.

**Synthesis of NAI-TPE-PyS**

A mixture solution of NAI-TPE-CHO (300 mg, 0.25 mmol), 3-(4-methylpyridin-1-ium-1-yl)propane-1-sulfonate (53 mg, 0.25mol), and piperidine catalyst (0.2 mL) was refluxed in 10 mL dry EtOH under $N_2$ for 48 hrs. The solution turned deep red. After cooling to room temperature, solvent was removed and the crude solid was purified by column with eluent of DCM: MeOH (10:1/v:v) to give red solid. $^1$H NMR (400 MHz, dmso) δ 9.21 (d, J = 6.9 Hz, 1H), 8.94 (d, J = 7.2 Hz, 1H), 8.47-8.41 (m, 10H), 8.29 (s, 2H), 8.20 (d, J = 7.0 Hz, 2H), 7.86-7.66 (m, 8H), 7.64 (d, J = 8.8 Hz, 1H), 7.51 (d, J = 8.5 Hz, 1H), 7.10 (dd, J = 14.4, 7.3 Hz, 4H), 7.03 – 6.95 (m, 6H), 6.84 (dd, J = 22.2, 10.3 Hz, 4H), 6.69 – 6.61 (m, 4H), 4.80 (t, J = 6.9 Hz, 2H), 4.20 (t, J = 7.0 Hz, 4H), 3.81 (t, J = 6.8 Hz, 4H), 2.45-2.41 (m, 2H), 2.22 (t, J = 6.9 Hz, 2H), 1.59-1.52 (m, 10H), 1.26-1.19 (m, 22H). $^{13}$C NMR (101 MHz, cdcl3) δ 164.12, 163.99, 133.75, 132.84, 131,10, 131.10, 127.57, 127.09, 126.87, 122,68, 56.55, 40.44, 29.68, 29.43, 29.30, 28.07, 27.09, 26.41, 22.66.

**Photoconductivity measurement and fabrication**

$SiO_2$/Si (2 μm thermal $SiO_2$) substrate was cleaned with acetone, IPA and D.I. water. A combination of photolithography, metal deposition, and lift-off were performed to pattern the electrical pad. Photoresist (AZ 5214) was spuncoat for 60 seconds at 3000 rpm, and a standard template mask with channel dimensions of 200 μm (length) x 5000 μm (width) was used as the electrode pattern. 5 nm Ti and 100nm Au was deposited using e-beam evaporation (Temescal, SL1800), and the residual photoresist was lifted-off, completing the electrode fabrication. A thin film of NAI-TPE-PyS was spun-coat on the device from a 6 wt% chloroform solution for 30 s with 500 rpm. Dark and light current measurements were performed. Photoconductivity of the device and I–t measurements were characterized by a Semiconductor Parameter Analyzer (Keysight B1500a). UV light source was provided by DYMAX LED DX-100.

## Author Contributions

Y.Z. synthesized molecule, performed chemical and photophysical characterizations, and prepared the manuscript. J.H. assisted in the synthesis and manuscript writing. P.J.G.S. performed theoretical calculations, contributed to the initial molecule design, and assisted in preparing the manuscript. H.U.C. performed photoconductivity measurements and analyzed data. S. D. performed Time-resolved photoluminescence (TRPL) measurements. A.M.A. and R.K. supervised the research and manuscript writing. All authors have given approval to the final version of the manuscript.

## Conflicts of interest

There are no conflicts to declare.

## Acknowledgements

The authors would like to thank the following researchers for insightful conversations: Dr. Jason Junge (USC Translational Imaging Center) and Yasaman Moradi. The authors acknowledge financial support from the Office of Naval Research (N00014-21-1-2048) and the Army Research Office (W911NF1810033).

## Notes and references

*Supplemental Information for:*

# Multifunctional photoresponsive organic molecule for electric field sensing and modulation

Yingmu Zhang[a], Jinghan He[b], Patrick J.G. Saris[a], Hyun Uk Chae[c], Subrata Das[c], Rehan Kapadia[c], Andrea M. Armani [a][b][c] *

**Table of Contents**

# 1 Synthesis of NAI-TPE-PyS

**Materials**

All commercially available chemical reagents and solvents including 1,8-naphthalic anhydride, ammonia, 1,10-dibromodecane 4-bromoiodobenzene, 4-Formylphenylboronic acid, bis(pinacolato)diboron, tetrakis(triphenylphosphine)palladium (Pd(PPh$_3$)$_4$), [1,1′-Bis(diphenylphosphino)ferrocene]dichloropalladium(II) (Pd(dppf)Cl$_2$), 4-Picoline, 1,3-propanesulton, piperidine, potassium carbonate (K$_2$CO$_3$), potassium acetate (KOAc), anhydrous sodium sulfate (Na$_2$SO$_4$), *N,N*-Dimethylformamide (DMF), dimethyl sulfoxide (DMSO), dichloromethane (DCM), chloroform, Tetrahydrofuran (THF), methanol (MeOH), ethanol (EtOH), ethyl acetate, hexane were either purchased from VWR or Sigma Aldrich. Unless otherwise noted, they were used without further purification. 1H-benz[de]isoquinoline-1,3(2H)-dione (1) and 4,4′-(2-(4-Bromophenyl)-2-phenylethene-1,1-diyl)- diphenol (5) were synthesized based on previously published procedures with modifications, as indicated.

$^1$H and $^{13}$C NMR spectra were recorded on a Varian Mercury 400 MHz spectrometer with 96-spinner sampler changer using either deuterated chloroform or DMSO as solvent, as indicated.

**Synthesis overview**

Scheme S1 shows the overview of the entire synthesis process for the NAI-TPE-PyS molecule. Compound 2, Compound 6, and Compound 8 are combined to form the desired product.

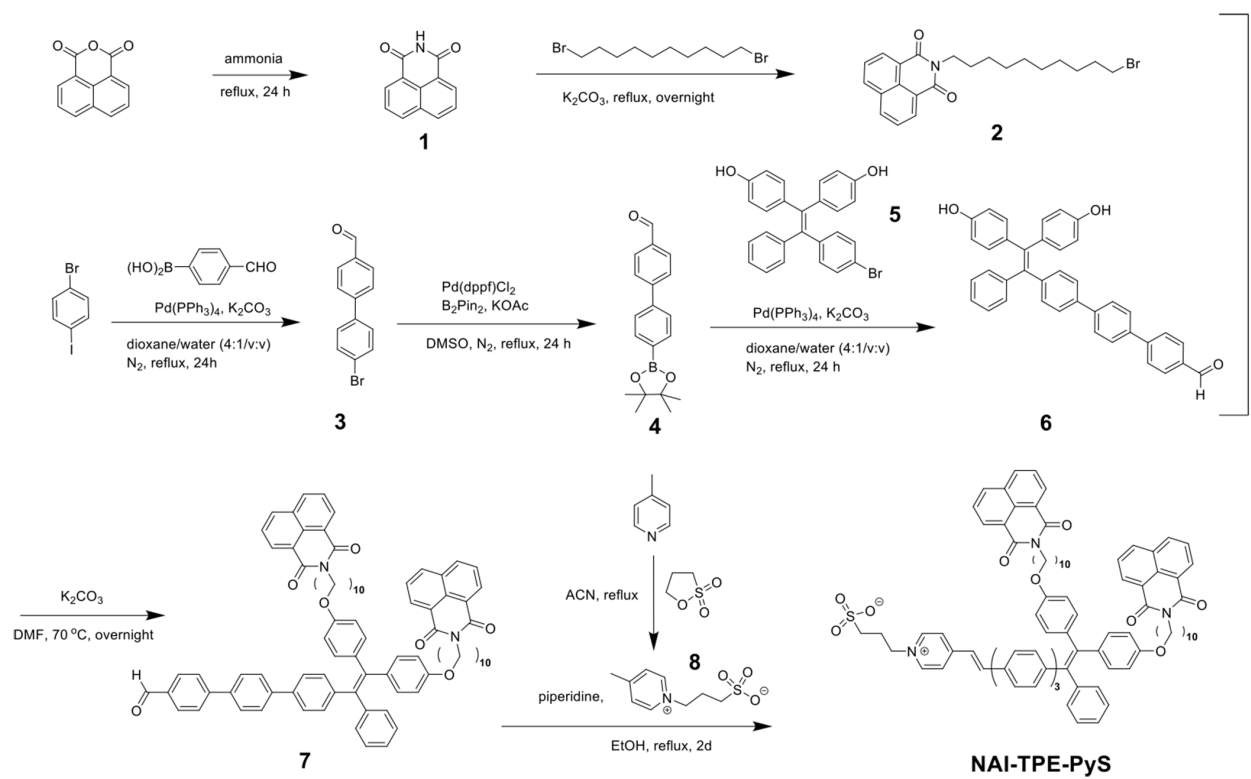

Scheme S1. Synthetic scheme of NAI-TPE-PyS.

**Synthesis of 2-(10-Bromodecyl)-1H-benz[de]isoquinoline-1,3(2H)-dione (Compound 2)**

Compound **1** was synthesized according to the previously published procedure.[1]

Into a 250 mL double-neck round-bottom flask was added **1** (2.0 g, 10.1 mmol) and dry *N,N*-Dimethylformamide (DMF, 80 mL). The mixture was stirred at 60°C for 12h to make **1** fully dissolved. Then, $K_2CO_3$ (2.79 g, 20.2 mmol) and 1,10-Dibromodecane (15.15 g, 50.5 mmol) were added, and the system was allowed to react at 60°C for another 24 h. After finishing, $K_2CO_3$ was removed by filtration, and the solvent was removed under vacuum. The crude product was further purified by silica column chromatography with ethyl acetate : hexane (1:9/v:v) as eluent to yield a white solid.

$^1$H NMR (400 Hz, Chloroform-*d*): δ 8.61 (d, J = 7.2 Hz, 2H), 8.21 (d, J = 8.1 Hz, 2H), 7.76 (t, 2H), 4.18 (t, 2H), 3.40 (t, J = 7.0 Hz, 2H), 1.84 (p, J=7.0 Hz, 2H), 1.74 (p, J=7.5 Hz, 3H), 1.47 = 1.24 (m, 12H). $^{13}$C-NMR (101MHz, Chloroform-d): δ: 164.20, 133.83, 131.58, 131.16, 128.16, 126.91, 122.77, 40.47, 34.04, 32.83, 29.38, 29.34, 29.27, 28.71, 28.15, 28.09, 27.09.

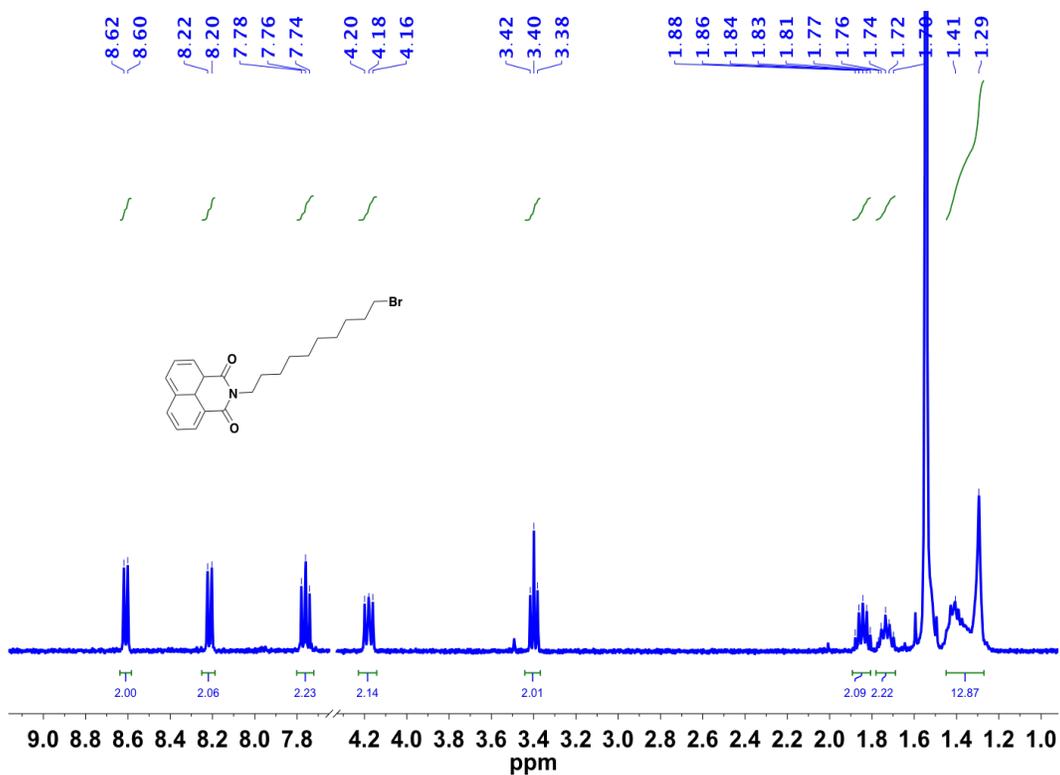

Fig. S1. $^1$H NMR of 2-(10-Bromodecyl)-1H-benz[de]isoquinoline-1,3(2H)-dione.

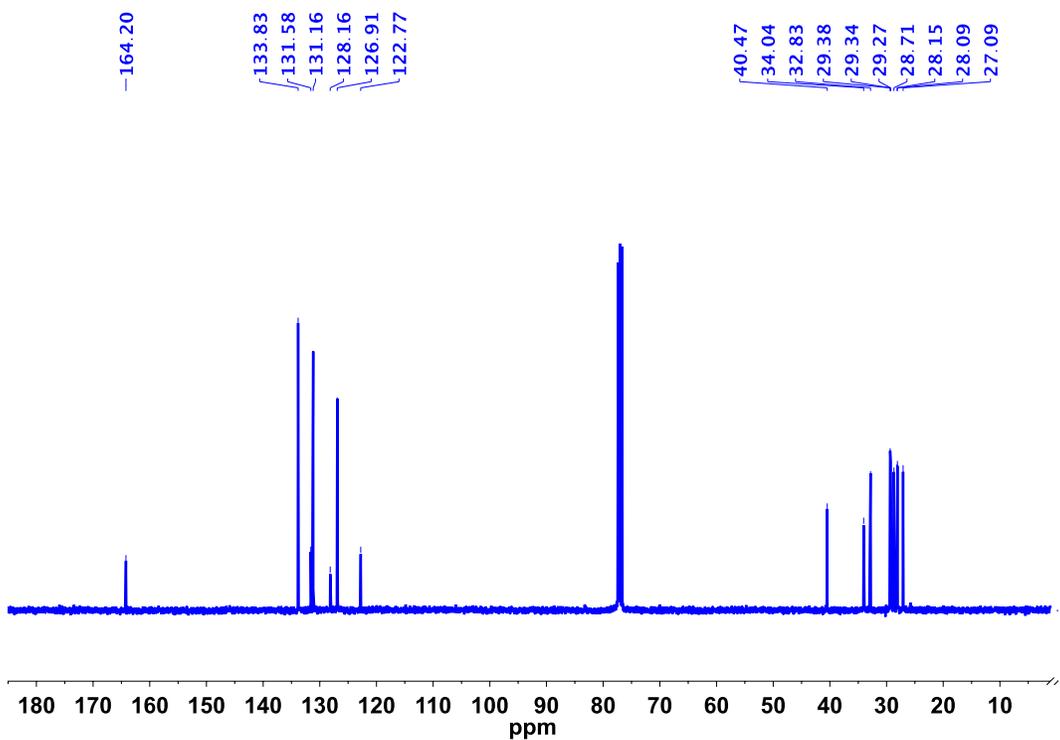

Fig. S2. $^{13}$C NMR of 2-(10-Bromodecyl)-1H-benz[de]isoquinoline-1,3(2H)-dione.

**Synthesis of 1-(4-Bromophenyl)-2,2-bis(4-hydroxyphenyl)-1-phenylethene (Compound 3)**

Into a 250 mL round-bottom flask were added 4-Bromoiodobenzene (20 mmol, 5.66 g), 4-Formylphenylboronic acid (20 mmol, 3.0 g), Pd(PPh$_3$)$_4$ (0.52 mmol, 600 mg) and K$_2$CO$_3$ (7.0g, 50 mmol). The flask was fitted on the Schlenk line, vacuum evacuated, and refilled with nitrogen alternately three times. A mixing solvent (dioxane/water: 80 mL/20 mL) was bubbled with nitrogen for 30 min and then transferred to the flask through a canula. The mixture was then allowed to react for 12 hours at 80 °C. After cooling to the room temperature, the mixture was poured into water and extracted with DCM three times. The combined organic part was dried with sodium sulfate, and the solvent was removed by vacuum. The obtained solid was then purified by column chromatography with eluent ethyl acetate : hexane (1: 10/v:v) to give a white powder.

$^1$H NMR (400 MHz, Chloroform-$d$) δ 10.07 (s, 1H), 7.95 (s, 2H), 7.71 (s, 2H), 7.62 (d, $J$ = 8.6 Hz, 2H), 7.51 (d, $J$ = 8.6 Hz, 2H). $^{13}$C-NMR (101MHz, Chloroform-d): δ 191.76, 145.88, 138.61, 135.45, 132.18, 130.35, 128.90, 127.49, 122.95.

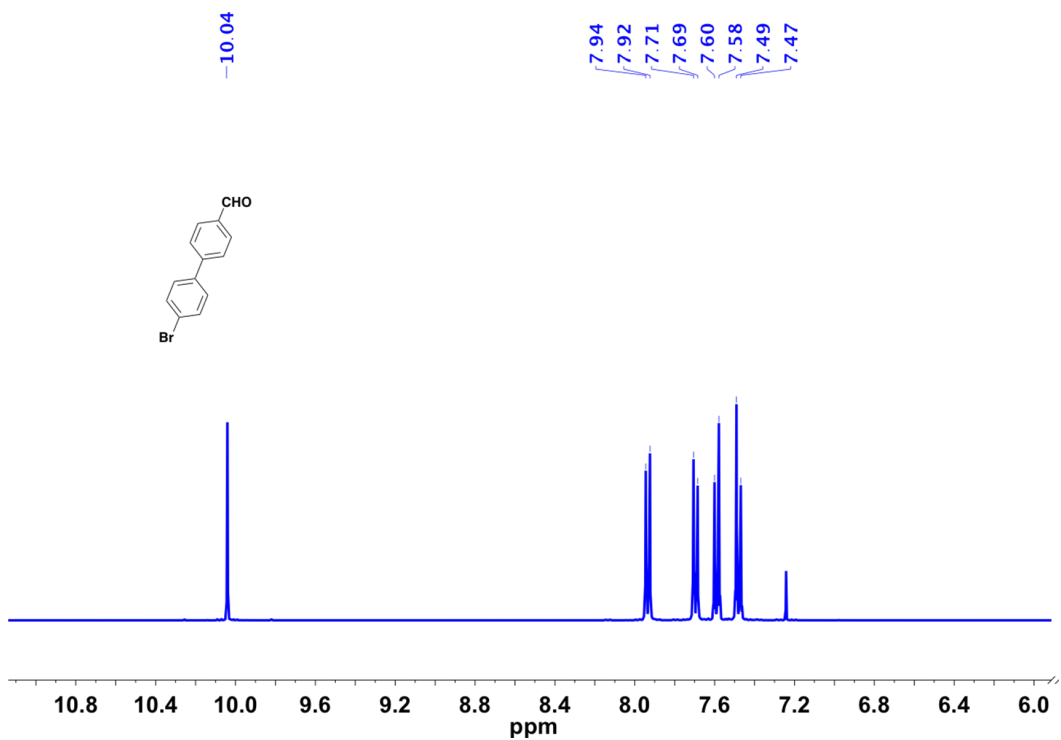

Fig. S3. $^1$H NMR of 1-(4-Bromophenyl)-2,2-bis(4-hydroxyphenyl)-1-phenylethene.

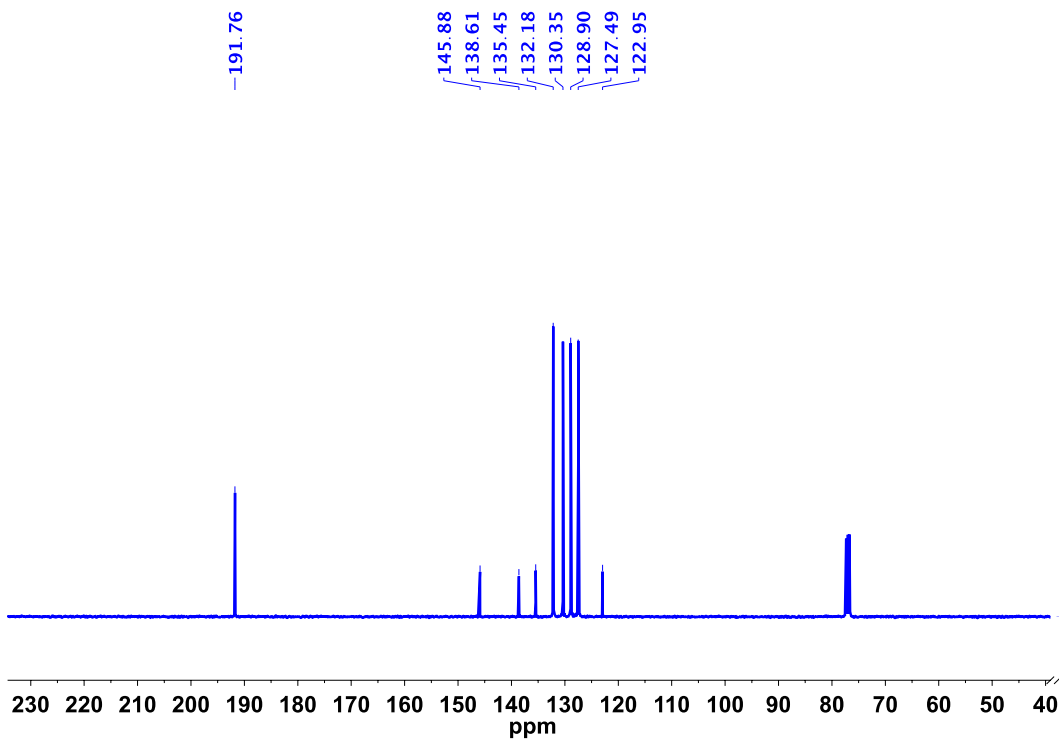

Fig. S4. $^{13}$C NMR of 1-(4-Bromophenyl)-2,2-bis(4-hydroxyphenyl)-1-phenylethene.

**Synthesis of 4'-(4,4,5,5-tetramethyl-1,3,2-dioxaborolan-2-yl)-[1,1'-biphenyl]-4-carbaldehyde (Compound 4)**

Compound **3** (1.0 g, 3.8 mmol), bis(pinacolato)diboron (1.5 g, 5.7 mmol), KOAc (1.2 g, 11.4 mmo) and Pd(dppf)Cl$_2$ (150 mg, 0.2 mmol) were added to a 250 mL Schleck flask with a stir bar. The flask was pumped under vacuum and refilled with N$_2$ three times before 50 mL degased DMSO was transferred to the system. The solution mixture was then heated at 100 $^0$C overnight under N$_2$. After cooling to room temperature, the mixture was poured into 200 mL of DI water, extracted with DCM twice and then washed with water three times. The combined organic layer was dried by anhydrous Na$_2$SO$_4$ and the organic solvent was pumped out. The crude product was then purified by column chromatography on silica gel with ethyl acetate/DCM (1 : 9/v:v) to give compound 4.

$^1$H NMR (400 MHz, Chloroform-*d*) δ 10.07 (s, 1H), 7.95 (s, 2H), 7.71 (s, 2H), 7.62 (d, *J* = 8.6 Hz, 2H), 7.51 (d, *J* = 8.6 Hz, 2H), 1.35 (s, 12 H). $^{13}$C-NMR (101MHz, Chloroform-d): δ 191.88, 146.97, 142.24, 135.42, 130.23, 127.79, 126.62, 83.97, 83.48, 25.01, 24.87.

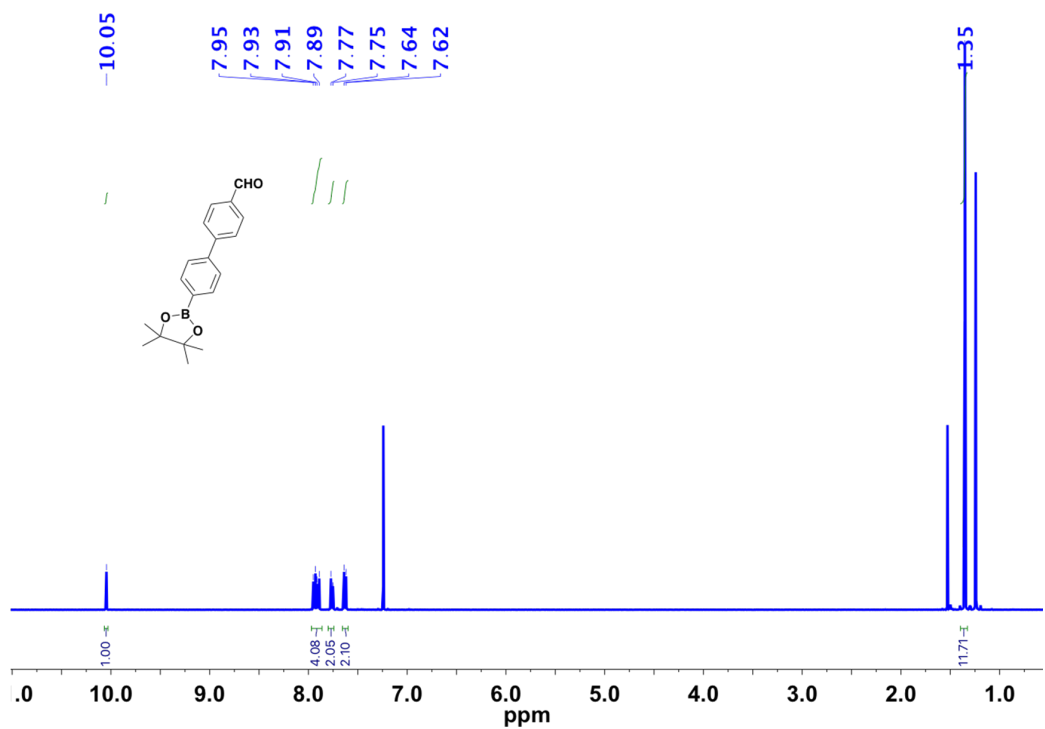

Fig. S5. ¹H NMR of 4'-(4,4,5,5-tetramethyl-1,3,2-dioxaborolan-2-yl)-[1,1'-biphenyl]-4-carbaldehyde.

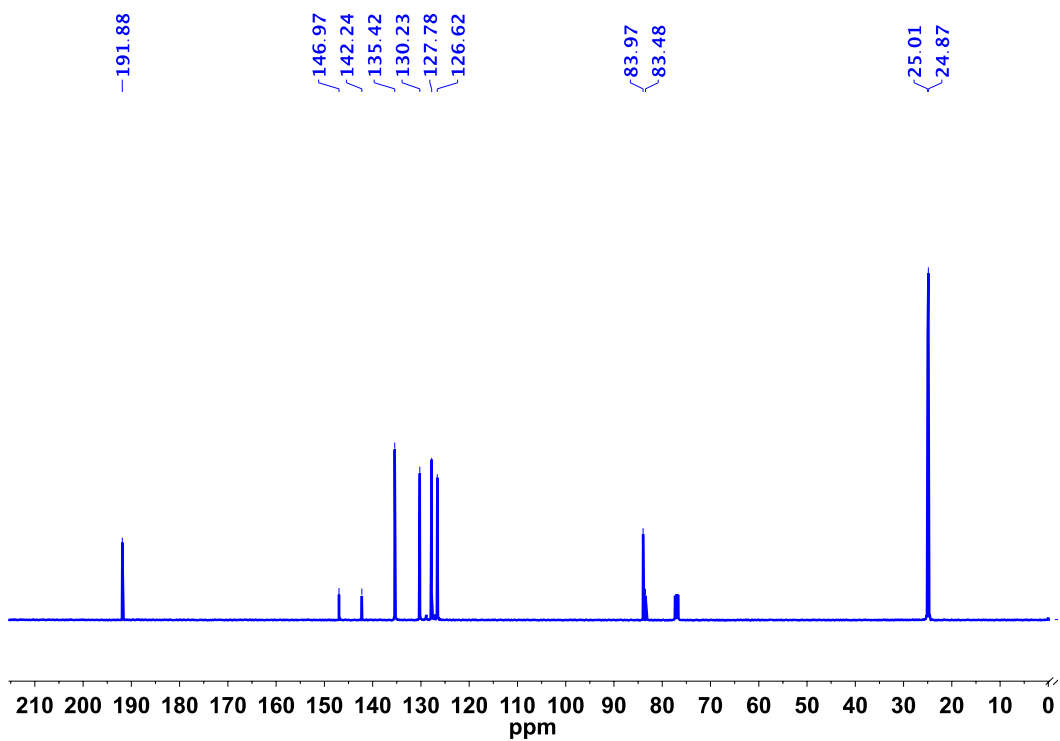

Fig. S6. ¹³C NMR of 4'-(4,4,5,5-tetramethyl-1,3,2-dioxaborolan-2-yl)-[1,1'-biphenyl]-4-carbaldehyde.

**Synthesis of 4''-(2,2-bis(4-hydroxyphenyl)-1-phenylvinyl)-[1,1':4',1''-terphenyl]-4-carbaldehyde (Compound 6)**

Compound **5** was synthesized according to previously reported literature.[2] Then, **4** (2.3 mmol, 0.7 g), **5** (1.9 mmol, 0.84 g), Pd(PPh$_3$)$_4$ (0.19 mmol, 200 mg) and K$_2$CO$_3$ (7.6 mmol, 1.05 g) were added into a 100 mL round-bottom flask. The flask was fitted on the Schlenk line, vacuumed and refilled with nitrogen alternately three times. A mixing solvent (dioxane/water: 40 mL/10 mL) was bubbled with nitrogen for 30 min and then transferred to the flask through a canula. The mixture was then allowed to react for 24 hours at 100 °C. After cooling to room temperature, the mixture was poured into DI water and the pH was adjusted to about 5. Then, the system was extracted with DCM and washed with water three times. The organic solvent was then removed and the crude solid was recrystallized by hexane/ DCM to give the pure product.

$^1$H NMR (400 MHz, Chloroform-*d*) δ 10.06 (s, 1H), 7.96 (d, *J* = 8.1 Hz, 2H), 7.79 (d, *J* = 8.4 Hz, 2H), 7.68 (s, 4H), 7.41 (d, *J* = 8.2 Hz, 3H), 7.12 (d, *J* = 2.2 Hz, 4H), 7.06 (s, 2H), 6.93 (dd, *J* = 16.6, 8.5 Hz, 4H), 6.59 (t, *J* = 9.0 Hz, 4H). $^{13}$C-NMR (101MHz, Chloroform-d): 191.91, 154.22, 154.14, 146.48, 144.10, 143.74, 140.85, 140.34, 138.79, 137.56, 136.40, 135.16, 132.78, 131.90, 131.42, 130.32, 127.75, 127.64, 127.46, 127.37, 126.22, 126.13, 114.68, 114.56.

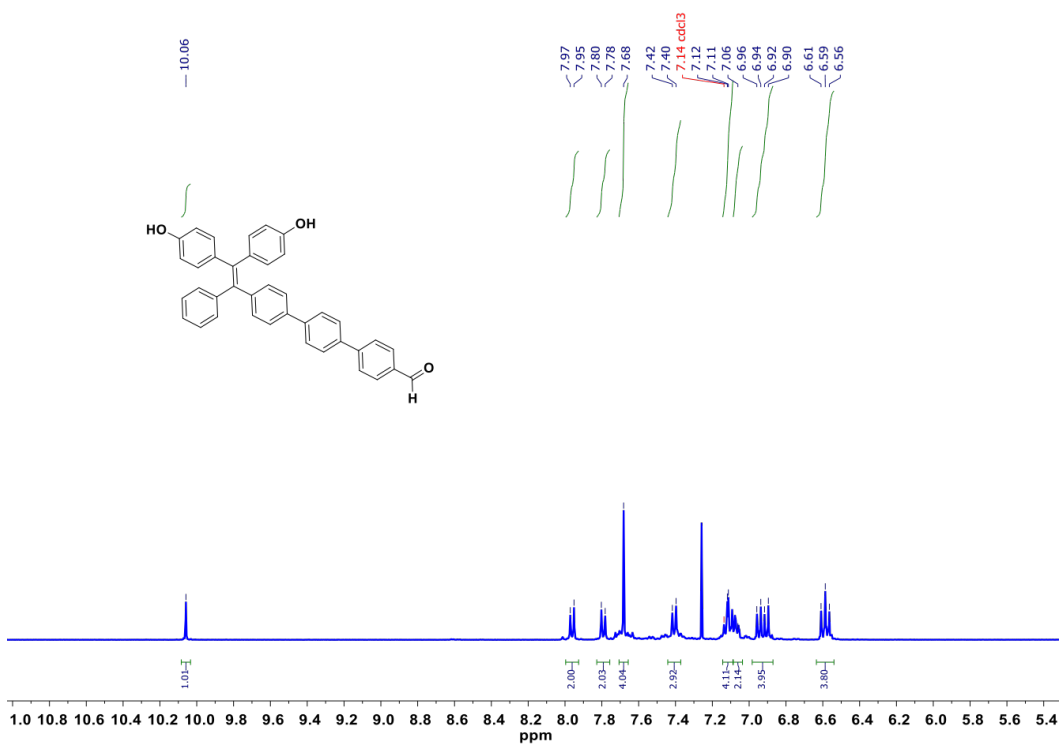

Fig. S7. $^1$H NMR of (Z)-4''-(1,2-bis(4-hydroxyphenyl)-2-phenylvinyl)-[1,1':4',1''-terphenyl]-4-carbaldehyde.

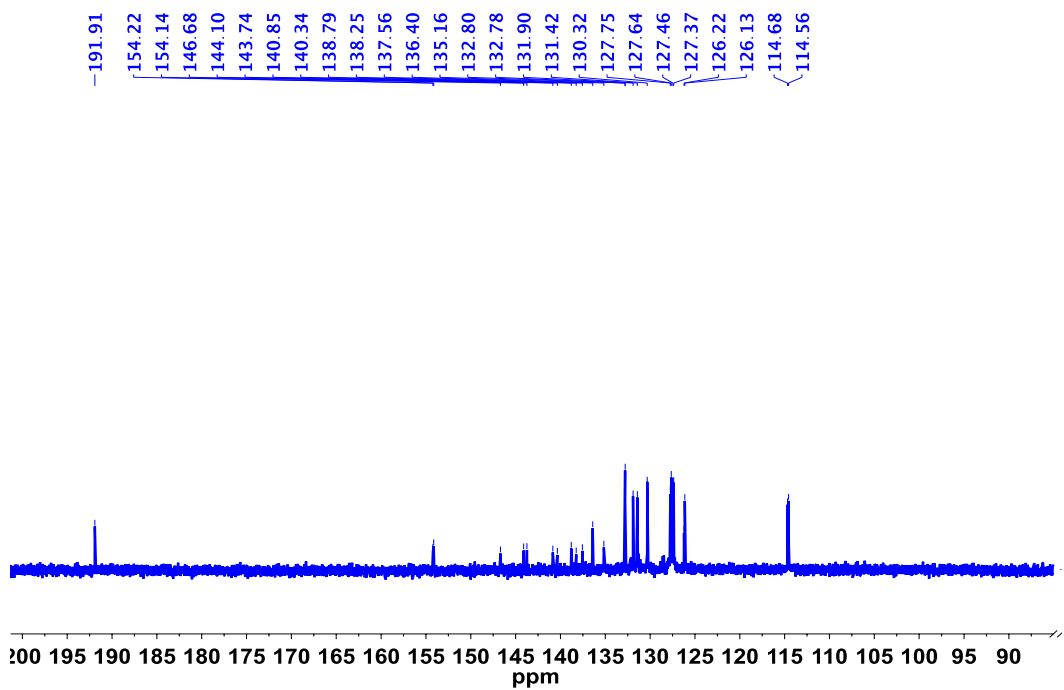

Fig. S8. $^{13}$C NMR of (Z)-4''-(1,2-bis(4-hydroxyphenyl)-2-phenylvinyl)-[1,1':4',1''-terphenyl]-4-carbaldehyde.

**Synthesis of 4''-(2,2-bis(4-((10-(1,3-dioxo-1H-benzo[de]isoquinolin-2(3H)-yl)decyl)oxy)phenyl)-1-phenylvinyl)-[1,1':4',1''-terphenyl]-4-carbaldehyde (NAI-TPE-CHO, Compound 7)**

Into a 250 mL two-necked round-bottom flask was added K$_2$CO$_3$ (248.8 mg, 1.8 mmol), **2** (500 mg, 1.2 mmol) and **6** (163 mg, 0.3 mmol). The flask was vacuumed and purged with dry N$_2$ three times. Then DMF (15 mL) was added and the reaction was stirred overnight at 70 °C. After cooling to room temperature, the mixture was poured into water, extracted with dichloromethane (DCM), washed with distilled water several times and dried with anhydrous magnesium sulfate. The crude product was purified by silica column chromatography with hexane and ethyl acetate (gradient to 1:1/v:v) as eluent to give **7** as a yellow viscous oil (200 mg, 56 %).

$^1$H NMR (400 MHz, Chloroform-d) δ 9.97 (s, 1H), 8.54 – 8.48 (m, 4H), 8.16 – 8.11 (m, 4H), 7.87 (d, J = 8.4 Hz, 2H), 7.74 – 7.63 (m, 7H), 7.60 (s, 4H), 7.32 (d, J = 8.4 Hz, 2H), 7.07 – 6.97 (m, 6H), 6.87 (dd, J = 17.3, 8.8 Hz, 4H), 6.60 – 6.51 (m, 4H), 4.12 – 4.07 (m, 4H), 3.79 (t, J = 6.5 Hz, 4H), 1.71 – 1.59 (m, 8H), 1.28 (d, J = 42.0 Hz, 24H). $^{13}$C-NMR (101MHz, Chloroform-d): 191.52, 164.19, 133.83, 132.61, 132.21, 131.56, 131.16, 130.30, 128.13, 127.79, 127.60,

127.42, 127.34, 126.09, 122.73, 113.90, 68.23, 40.48, 32.75, 32.64, 29.38, 29.32, 29.29, 29.28, 28.08, 28.04, 27.09, 26.02, 25.95, 25.78, 25.64.

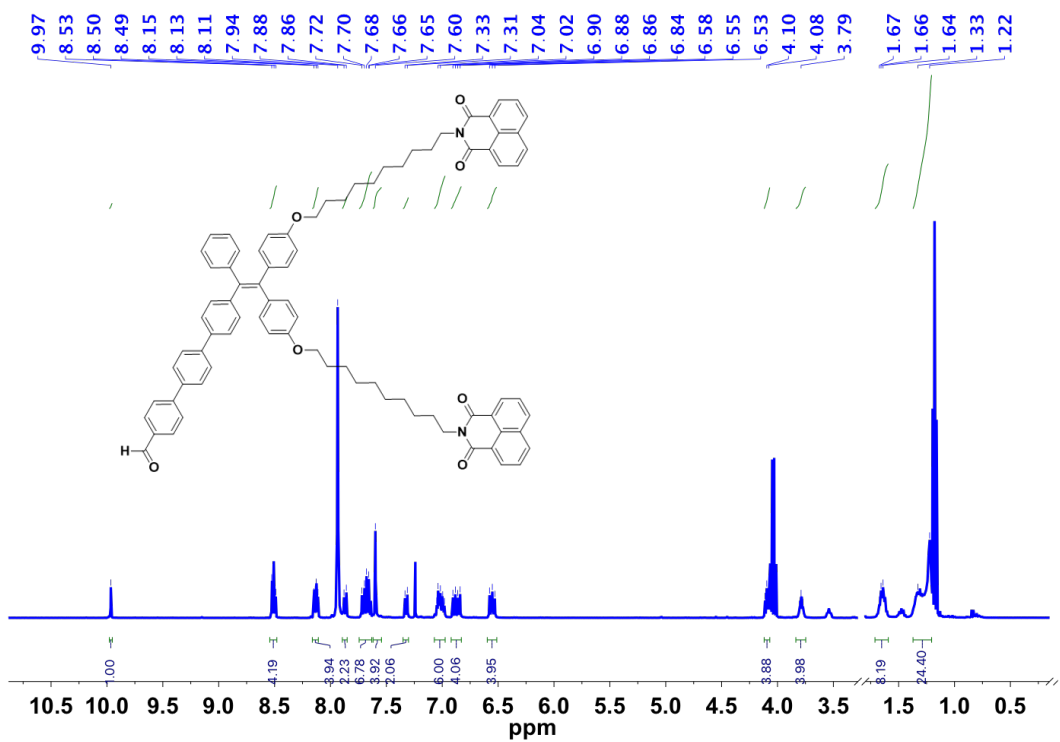

Fig. S9. ¹H NMR of NAI-TPE-CHO.

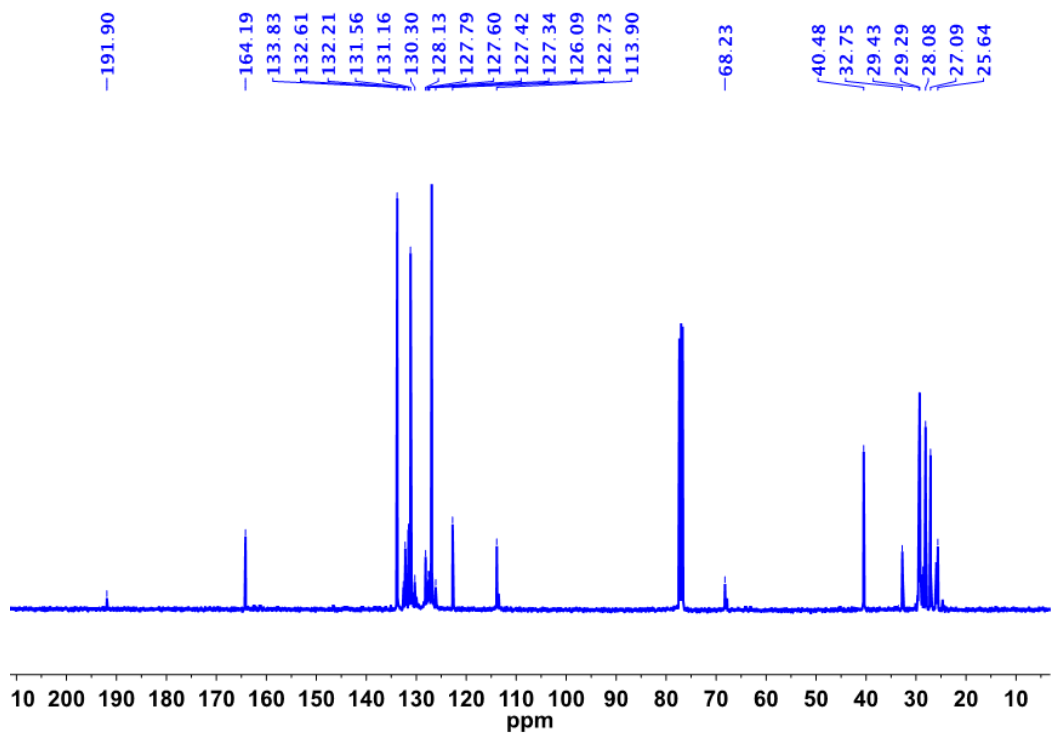

Fig. S10. $^{13}$C NMR of NAI-TPE-CHO.

**Synthesis of 3-(4-methylpyridin-1-ium-1-yl)propane-1-sulfonate (Compound 8)**

4-Picoline (1.8 g, 20 mmol) was dissolved in 15 ml MeCN. 1,3-propanesultone (3.7 g, 30 mmol) was added and the reaction mixture was heated to 80 °C for 4 hours. After completion, the crude product was precipitated. The solid was filtered and washed with ethyl ether to give compound **8** (86%) as a white solid.

$^1$H NMR (400 MHz, DMSO-d6) δ 8.88 (d, J = 6.8 Hz, 2H), 7.94 (d, J = 6.3 Hz, 2H), 4.63 (t, J = 6.9 Hz, 2H), 2.56 (s, 3H), 2.36 (t, J = 7.2 Hz, 2H), 2.17 (q, J = 6.9 Hz, 2H). $^{13}$C NMR (101 MHz, DMSO-d$_6$) δ 159.21, 144.37, 128.78, 59.22, 47.38, 27.69, 21.82.

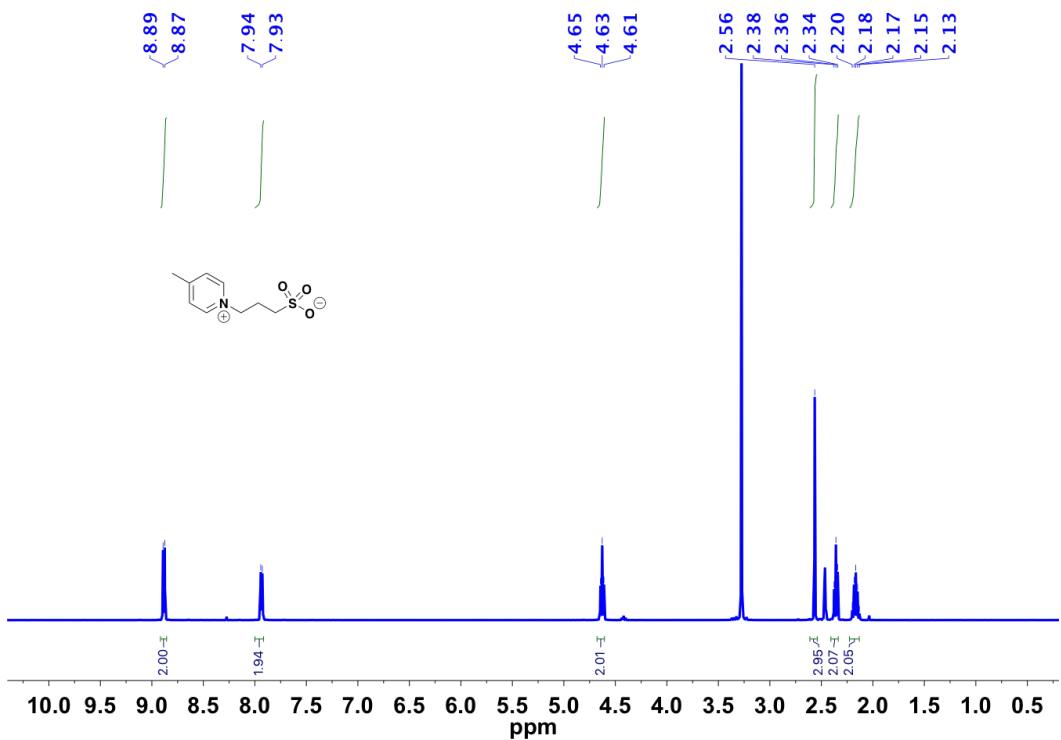

Fig. S11. ¹H NMR of 3-(4-methylpyridin-1-ium-1-yl)propane-1-sulfonate.

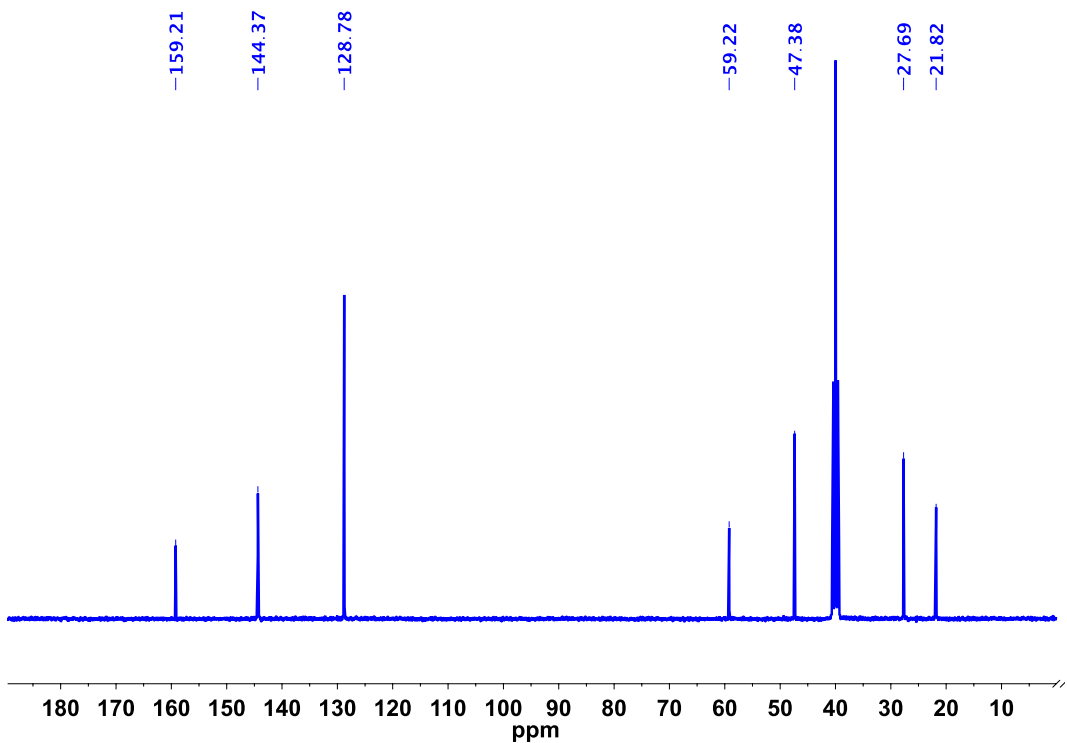

Fig. S12. ¹³C NMR of 3-(4-methylpyridin-1-ium-1-yl)propane-1-sulfonate.

**Synthesis of (E)-3-(4-(2-(4''-(2,2-bis(4-((10-(1,3-dioxo-1H-benzo[de]isoquinolin-2(3H)-yl)decyl)oxy)phenyl)-1-phenylvinyl)-[1,1':4',1''-terphenyl]-4-yl)vinyl)pyridin-1-ium-1-yl)propane-1-sulfonate (NAI-TPE-PyS)**

A mixture solution of **7** (300 mg, 0.25 mmol), **8** (53 mg, 0.25mol), and piperidine catalyst (0.2 mL) was refluxed in 10 mL dry EtOH under $N_2$ for 48 hrs. The solution turned deep red. After cooling to room temperature, the solvent was removed and the crude solid was purified by column with eluent of DCM: MeOH (10:1/v:v) to give a red solid (83 %).

$^1$H NMR (400 MHz, dmso) δ 9.21 (d, $J$ = 6.9 Hz, 1H), 8.94 (d, $J$ = 7.2 Hz, 1H), 8.47-8.41 (m, 10H), 8.29 (s, 2H), 8.20 (d, $J$ = 7.0 Hz, 2H), 7.86-7.66 (m, 8H), 7.64 (d, $J$ = 8.8 Hz, 1H), 7.51 (d, $J$ = 8.5 Hz, 1H), 7.10 (dd, $J$ = 14.4, 7.3 Hz, 4H), 7.03 – 6.95 (m, 6H), 6.84 (dd, $J$ = 22.2, 10.3 Hz, 4H), 6.69 – 6.61 (m, 4H), 4.80 (t, $J$ = 6.9 Hz, 2H), 4.20 (t, $J$ = 7.0 Hz, 4H), 3.81 (t, $J$ = 6.8 Hz, 4H), 2.45-2.41 (m, 2H), 2.22 (t, $J$ = 6.9 Hz, 2H), 1.59-1.52 (m, 10H), 1.26-1.19 (m, 22H). $^{13}$C NMR (101 MHz, cdcl$_3$) δ 164.12, 163.99, 133.75, 132.84, 131,10, 131.10, 127.57, 127.09, 126.87, 122,68, 56.55, 40.44, 29.68, 29.43, 29.30, 28.07, 27.09, 26.41, 22.66

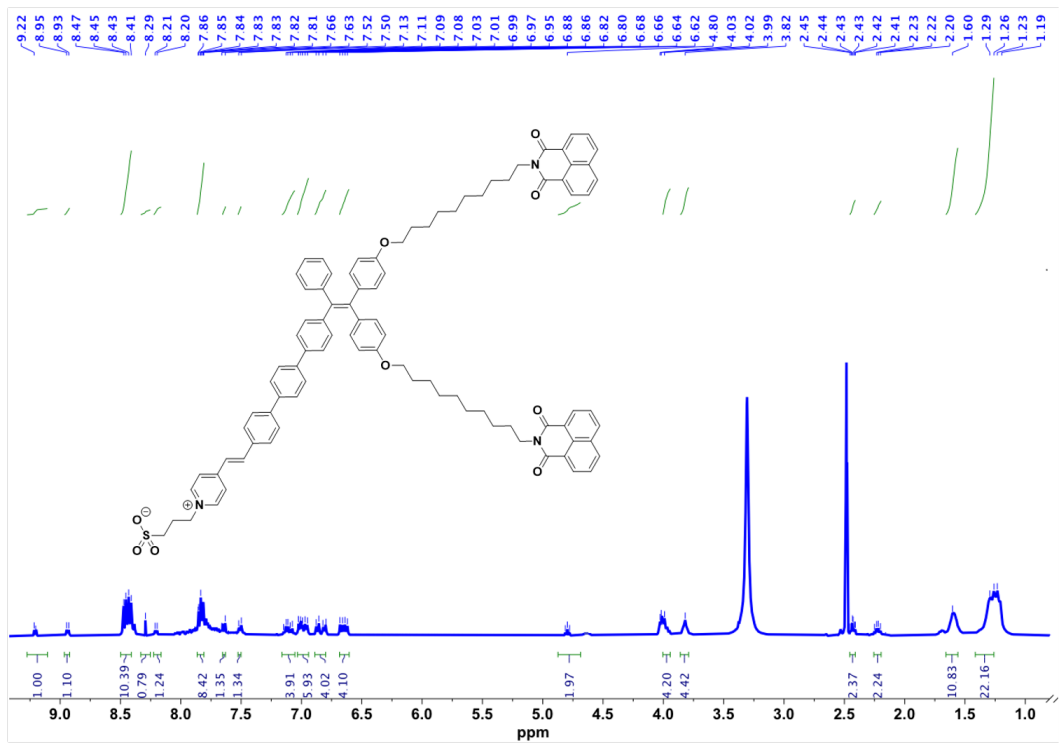

Fig. S13. $^1$H NMR of NAI-TPE-PyS.

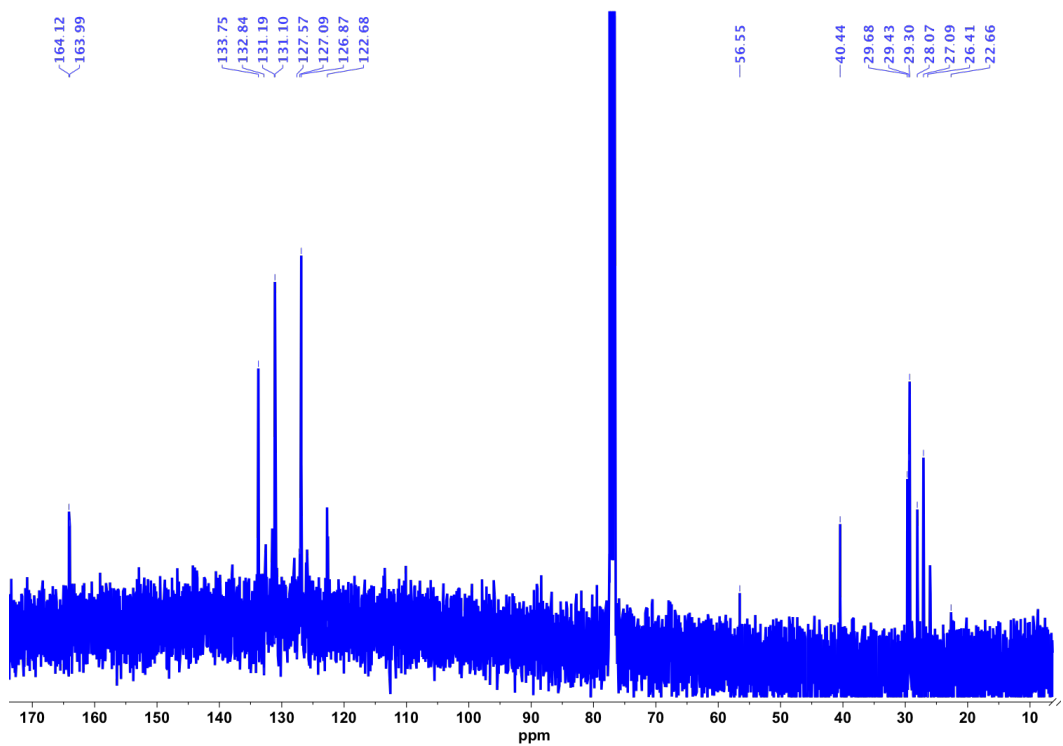

Fig. S14. $^{13}$C NMR of NAI-TPE-PyS.

## 2 Optical Behavior and Photophysical Properties

Dulbecco's phosphate-buffered saline (DPBS, 1X) was purchased from Thermo Fisher Scientific. Selective solvents *N*,*N*-Dimethylformamide (DMF), dimethyl sulfoxide (DMSO), dichloromethane (DCM), chloroform, Tetrahydrofuran (THF), methanol (MeOH), ethanol (EtOH) were either purchased from VWR or Sigma Aldrich.

All UV-Vis absorption spectra were measured on Beckman Coulter Life Science UV/Vis spectrophotometer, DU 730 with wavelength resolution of 2 nm. Steady-state fluorescence spectra were recorded on Horiba Scientific Fluoromax-4 spectrofluorometer with excitation slit width of 5 nm and emission slit width of 5 nm. Quantum yield was determined by a Quanta-φ integrating sphere. Powder sample and aqueous solution of the sample (10 μM, containing 1% DMSO) were prepared for the measurements.

The optical absorption, steady-state emission, and quantum yield were measured under several different conditions to understand the role of pH, solvent polarity, and fluorophore concentration on emission intensity and emission wavelength. Measurements were performed on both solution and thin film samples. Solution samples in selective solvents were prepared by serial dilution to give the desired concentrations, and thin film samples were prepared by dropping DCM solution of NAI-TPE-PyS on Swiss glass slide (size: 25 mm x 75 mm, thick: 1.0 mm) followed by air-dry evaporation.

Fig. S15 shows the UV-Vis absorption measurements of several synthetic precursors, the final product both in DCM and in PBS, and the final product deposited as a thin film on quartz substrate.

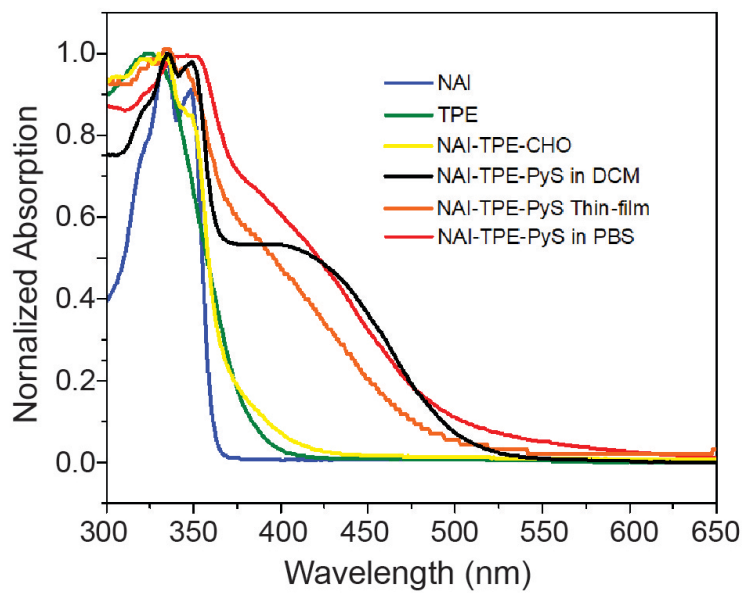

Fig. S15. Absorption Comparison among NAI-TPE-PyS and synthetic precursors in solvents and solid state.

Fig. S16 and Fig. S17 shows the results from a study of the UV-Vis absorption behavior and of the fluorescence behavior of the final product in a range of solvents. The maximum absorption and emission wavelengths for each solvent is summarized in Table S1. Optical images of each solution are shown in Fig. S18.

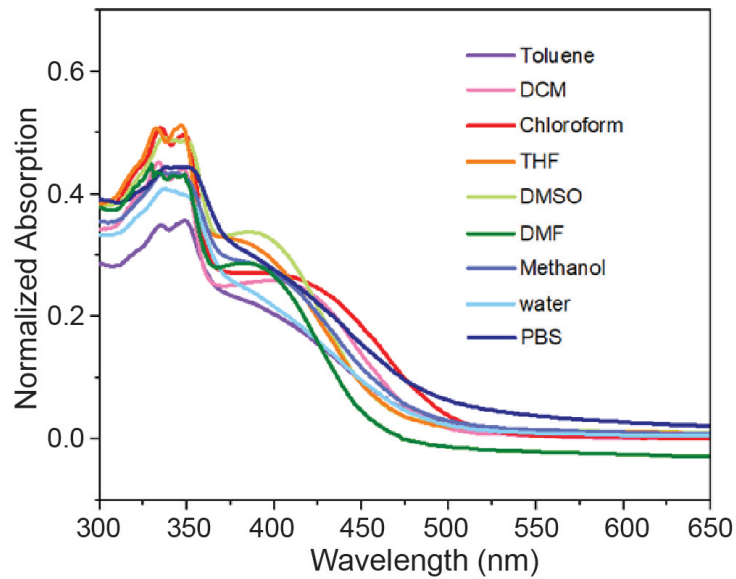

Fig. S16. Absorption spectra of NAI-TPE-PyS in different solvents. Concentration: 10 μM.

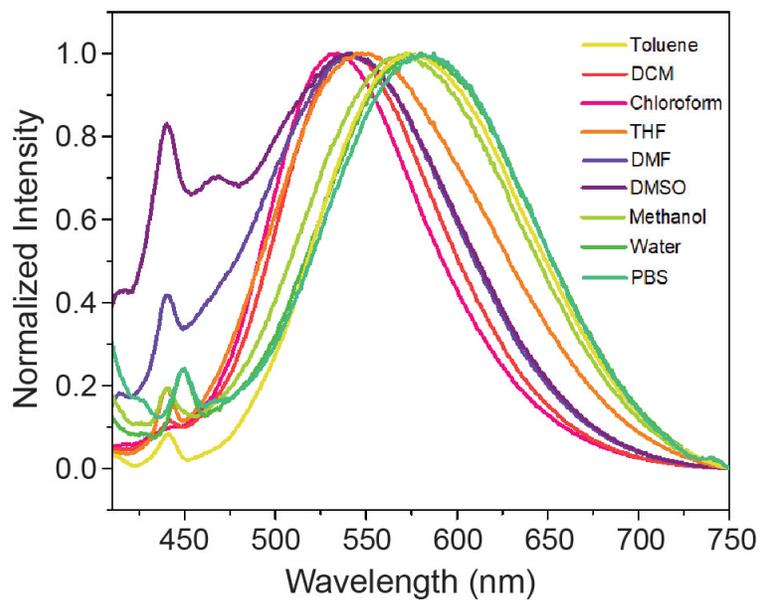

Fig. S17. Emission spectra of NAI-TPE-PyS in different solvents. Concentration: 10 μM, $\lambda_{ex}$= 390 nm.

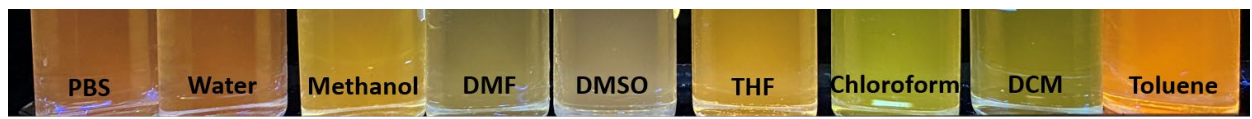

Fig. S18. Photographic images of NAI-TPE-PyS emission in different solvents at $\lambda_{ex}$= 365 nm.

Table S1 summarizes the detailed photophysical data of NAI-TPE-PyS in different solvents including orientation polarizability of selective solvents ($\Delta f$), $\lambda_{abs}$ and $\lambda_{em}$ derived from the UV and FL peaks and calculated $\nu_{abs}$, $\nu_{em}$ and Stoke shift ($\nu_{abs}-\nu_{em}$). Figure 19 depicts the Lippert-Mataga relation by plotting Stoke shifts against solvent orientation polarity, showing a positive solvotochromic effect.

The relationship between the Stoke shift ($\nu_{abs}-\nu_{em}$) of the fluorophore and orientation polarizability $f(\varepsilon, n)$ can be described by the Lippert-Mataga equation:

$$hc(\nu_{abs}-\nu_{em}) = hc(\nu_{abs}^0-\nu_{em}^0) + \frac{2(\mu_e-\mu_g)^2}{a^3} f(\varepsilon, n)$$

where $h$ is Plank's constant, $c$ is the velocity of light, $f$ is the orientational polarizability of the solvent, $\nu_{abs}^0-\nu_{em}^0$ corresponds to the Stokes shifts when $f$ is zero, $\mu_e$ is the excited-state dipole moment, $\mu_g$ is the ground-state dipole moment, a is the solvent Onsager cavity radius derived from Avogadro number (N), molecular weight (M) and density (d =1.0 g/cm³), and ε and n are the solvent dielectric and the solvent refractive index, respectively.

$\mu_e$ can be calculated according to the equation:

$$\mu_e = \mu_g + \left\{ \frac{hca^3}{2} * \left[ \frac{d(\nu_{abs}-\nu_{em})}{df(\varepsilon, n)} \right] \right\}^{\frac{1}{2}}$$

where $\mu_g$ was estimated around 29.3 D from DFT in the gas phase at the B3LYP/6-31g* level of theory 31G(d) level, and $\mu_e$ was calculated to be 59.1 D.

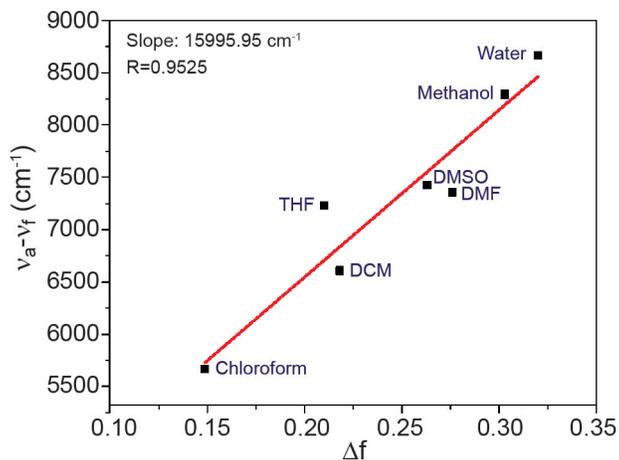

Figure S19. Lippert-Mataga plot for NAI-TPE-PyS in different solvents as a function of solvent polarity. ($\Delta f$: orientation polarizability; $\nu_{abs}-\nu_{em}$: Stoke shifts).

Table S1. Detailed photophysical data of NAI-TPE-PyS in selective solvents.

| Solvent | $\Delta f$ | $\lambda_{abs}$ (nm) | $\lambda_{em}$ (nm) | $\nu_{abs}$ (cm$^{-1}$) | $\nu_{em}$ (cm$^{-1}$) | $\nu_{abs}-\nu_{em}$ (cm$^{-1}$) |
|---|---|---|---|---|---|---|
| DCM | 0.218 | 398 | 540 | 25125.63 | 18518.52 | 6607.11 |
| chloroform | 0.149 | 410 | 534 | 24390.24 | 18726.59 | 5663.652 |
| THF | 0.210 | 391 | 545 | 25575.45 | 18348.62 | 7226.824 |
| DMSO | 0.263 | 386 | 541 | 25906.74 | 18484.29 | 7422.447 |
| DMF | 0.276 | 387 | 541 | 25839.79 | 18484.29 | 7355.505 |
| Methanol | 0.309 | 388 | 541 | 25773.2 | 17482.52 | 8290.678 |
| water | 0.32 | 386 | 580 | 25906.74 | 17241.38 | 8665.356 |
| PBS | – | 398 | 582 | 25125.63 | 17182.13 | 6607.11 |

## 3  Aggregation Behaviors

The AIE feature of the compound was studied by adding anti-solvent THF into the DMSO solution of NAI-TPE-PyS (20 µM) gradually with volume fraction ranging from 0% to 99 %. The measurement of AIE property requires an environment where the compound form aggregates or micelles. As NAI-TPE-PyS is well dispersed in DMSO and shows low solubility in THF, the mixture DMSO/THF are chosen to provide an environment for NAI-TPE-PyS to form aggregates.

The emission intensity of the mixtures was then measured at λex= 390 nm. Figure S20 (a) shows the PL spectra of NAI-TPE-PyS in DMSO/THF mixtures with different THF fractions ($f_{THF}$), and Fig. S20(b) indicates the corresponding relative PL intensity.

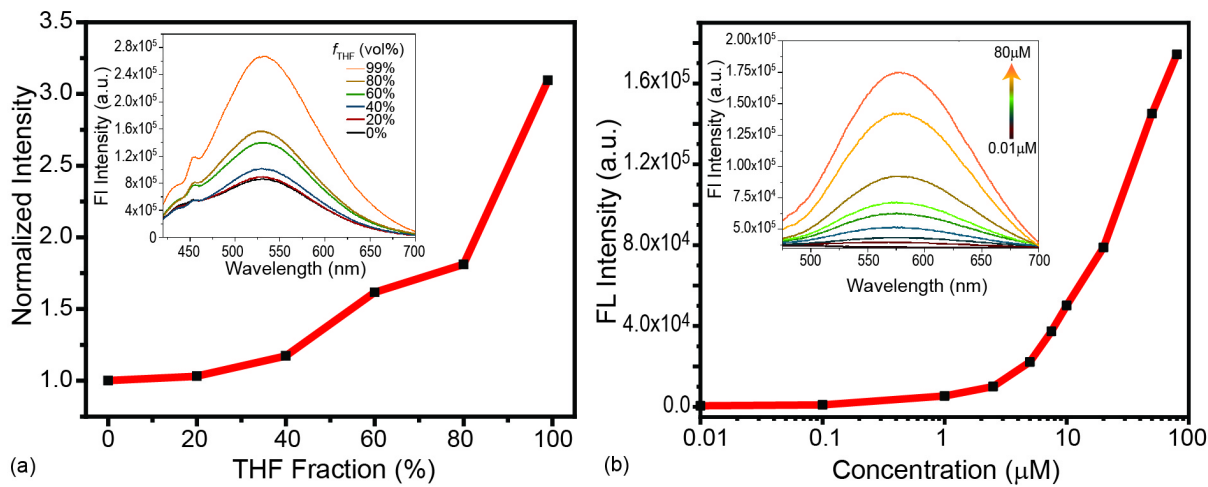

Fig. S20. AIE behavior of NAI-TPE-PyS. (a) Plot of relative PL intensity versus THF fraction. Inset: PL spectra of NAI-TPE-PyS in DMSO/THF mixtures with different THF fractions ($f_{THF}$). Concentration of NAI-TPE-PyS is 20 μM; $λ_{ex}$= 390 nm. (b) Plot of PL intensity versus NAI-TPE-PyS concentration in water. Inset: PL spectra of aqueous solutions of NAI-TPE-PyS at concentrations ranging from 0.01μM to 80μM ($λ_{ex}$= 390 nm).

DLS is used to prove the formation of NAI-TPE-PyS aggregates in aqueous media. DMSO solution of NAI-TPE-PyS (8 mM) was diluted with DI water to give 50 μM sample solution in DMSO/water mixture with water fraction of 99 % (v%). The measurements were conducted on cuvette-based DLS instrument DynaPro®NanoStar®, WYATT Technology. The obtained particle size distribution was plotted in Fig. S21 and particles with effective diameter of 500 nm were given by DYNAMICS® software.

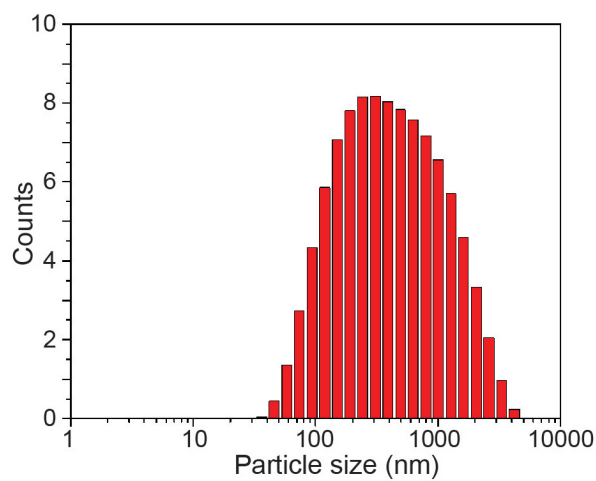

Figure S21. Particle size distribution of NAI-TPE-PyS aggregates in DMSO/THF mixture with a 99% THF fraction. Concentration: 50 μM.

## 4 Two-photon excitation

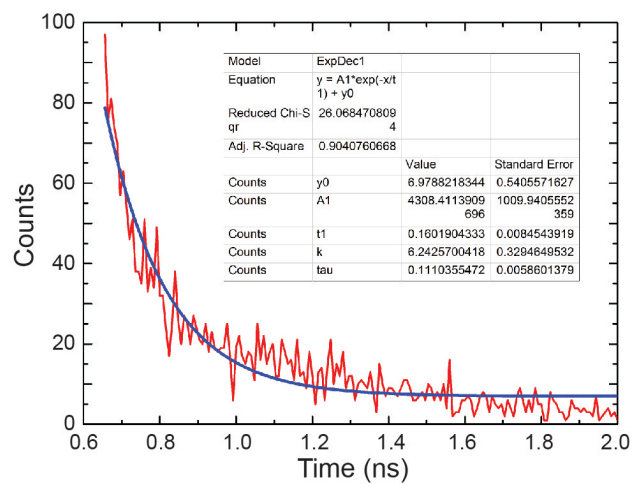

Fig. S22. TCSPC histogram of NAI-TPE-PyS under 750 nm excitation. Red: transient decay trace; Blue: exponential fit line; Inset: exponential fit parameters.

Time-resolved photoluminescence (TRPL) measurements were performed at room temperature using a standard confocal microscope-based Time-Correlated Single Photon

Counting (TCSPC) setup. For this measurement, the sample was dropped cast on a glass slide and a 750 nm pulsed laser (Spectra-Physics, Mai Tai, Mode-Locked Ti:Sapphire Laser) with a pulse width of less than 200 fs and repetition rate of 80 MHz was focused on the sample with a microscope objective. The PL signal from the sample was then passed through two short pass filters with a cut-off wavelength of 550 nm in order to reject the laser light from the laser radiation. Finally, the PL photons were detected with a Si Avalanche Photo Diode (Si-APD) and the PL transient was measured as a function of time. The results are shown in Fig. S22.

## 5  Photoconductivity

$SiO_2$/Si (2 μm thermal $SiO_2$) substrate was cleaned with acetone, IPA and D.I. water. A combination of photolithography, metal deposition, and lift-off were performed to pattern the electrical pad. Photoresist (AZ 5214) was spuncoat for 60 seconds at 3000 rpm, and a standard template mask with channel dimensions of 200 μm (length) x 5000 μm (width) was used as the electrode pattern. 5 nm Ti and 100nm Au was deposited using e-beam evaporation (Temescal, SL1800), and the residual photoresist was lifted-off, completing the electrode fabrication. A thin film of NAI-TPE-PyS was spun-coat on the device from a 6 wt% chloroform solution for 30 s with 500 rpm. Dark and light current measurements were performed by measuring by the voltage across the electrical contact pad. Photoconductivity of the device and I−t measurements were characterized by a Semiconductor Parameter Analyzer (Keysight B1500a). UV light source was provided by DYMAX LED DX-100 with $\lambda_{ex}$=350 nm.

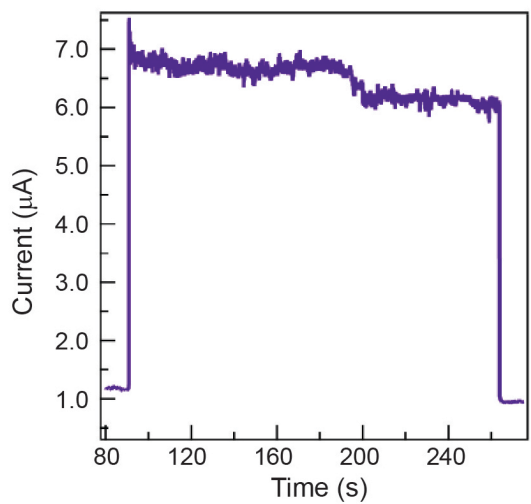

Figure S23. I-t curve of the device at fixed voltage of 10 V upon light illumination with 19 mW optical power.

Table S2. The responsivity studies of our molecules.

| Wavelength (nm) | Incident Optical Power (mW) | Responsivity (µA/W) |
| --- | --- | --- |
| dark | N/A | 0.07 |
| 350 | 8 | 0.16 |
| 350 | 12.5 | 0.22 |
| 350 | 16 | 0.30 |
| 350 | 19 | 0.41 |